\documentclass[12pt]{article}

\usepackage{amsmath, amsfonts, amssymb}
\usepackage{amsthm}

\allowdisplaybreaks

\usepackage{microtype}
\usepackage{dsfont}
\usepackage{appendix}
\usepackage{enumitem}

\usepackage{algorithm}
\usepackage{algpseudocodex}

\algrenewcommand{\algorithmicrequire}{\textbf{Input:}}
\algrenewcommand{\algorithmicensure}{\textbf{Output:}}

\usepackage{natbib}

\usepackage{xcolor}
\usepackage{graphicx}
\usepackage{float}
\usepackage{subfigure}
\usepackage{multirow}
\usepackage{booktabs} 

\usepackage{geometry}
\usepackage{titlesec}

\titleformat{\paragraph}
{\normalfont\normalsize\bfseries}{\theparagraph}{1em}{}
\titlespacing*{\paragraph}
{0pt}{3.25ex plus 1ex minus .2ex}{1.5ex plus .2ex}

\usepackage{hyperref}
\hypersetup{
            colorlinks={true}, 
            linkcolor={blue}, 
            citecolor=blue,
            pdfencoding=auto, 
            psdextra
            }
            
\usepackage[capitalise,compress]{cleveref} 

\numberwithin{equation}{section}

\newtheorem{assumption}{Assumption}[section]
\newtheorem{lemma}{Lemma}[section]
\newtheorem{theorem}{Theorem}[section]
\newtheorem{proposition}{Proposition}[section]
\newtheorem{remark}{Remark}[section]

\newtheorem{example}{Example}[section]

\Crefname{assumption}{Assumption}{Assumptions}
\Crefname{definition}{Definition}{Definitions}
\Crefname{lemma}{Lemma}{Lemmas}

\newcommand{\E}{\mathbb{E}}
\newcommand{\Pp}{\mathbb{P}}
\newcommand{\Var}{\operatorname{Var}}
\newcommand{\Cov}{\operatorname{Cov}}

\newcommand{\op}{o_{\Pp}}
\newcommand{\Op}{O_{\Pp}}
\newcommand{\R}{\mathbb{R}}

\newcommand{\G}{\mathbb{G}}
\newcommand{\Pn}{\mathbb{P}_{n}}
\newcommand{\QN}{\mathbb{Q}_{N}}

\newcommand{\thetahat}{\widehat\theta_{\mathrm{DA}}}
\newcommand{\thetabc}{\widehat\theta_{\mathrm{DDA}}}

\newlist{assumpenum}{enumerate}{1} 
\setlist[assumpenum]{label=(\roman*), ref=\theassumption~(\roman*)}
\crefalias{assumpenumi}{assumption} 

\newlist{lemmaenum}{enumerate}{1} 
\setlist[lemmaenum]{label=(\roman*), ref=\thelemma~(\roman*)}
\crefalias{lemmaenumi}{lemma} 

\newlist{propenum}{enumerate}{1} 
\setlist[propenum]{label=(\roman*), ref=\theproposition~(\roman*)}
\crefalias{propenumi}{proposition} 

\DeclareMathOperator{\argmin}{\mathop{\mathrm{argmin}}}

\makeatother
\title{Distribution-Aligned \(M\)-Estimation}

\author{Yanqin Fan\thanks{Department of Economics, University of Washington. Email: fany88@uw.edu}, Ruixuan Liu\thanks{CUHK Business School, Chinese University of Hong Kong, Email:ruixuanliu@cuhk.edu.hk}, Carlos A. Manzanares\thanks{Amazon. Email: manzcarl@amazon.com}, and Hyeonseok Park\thanks{Institute for Advanced Economic Research, Dongbei University of Finance and Economics. Email: hynskpark21@dufe.edu.cn} }

\date{\today}

\begin{document}

\maketitle

\begingroup
\renewcommand{\thefootnote}{}
\footnotetext[0]{ We used ChatGPT, Codex and Claude for coding, proof checking, and feedback. All remaining errors are our own.}
\endgroup
\begin{abstract}
    We propose a distribution-aligned approach to M-estimation when externally generated surrogates are available. Our method provides a unified view of prediction-powered inference (PPI) and agentic economic modeling (AEM) through a common object: the conditional distribution of the true outcome given covariates and the surrogate. This perspective also connects our setting to the classical literature on missing data and measurement error with validation data. Using labeled observations, we estimate this conditional distribution and then impute missing outcomes in unlabeled data by integrating the original loss with respect to the fitted conditional distribution. This plug-in distributional alignment (DA) preserves the convex structure of the original loss and attains the same efficiency as the current benchmark in PPI. The asymptotic representation of DA relies on the standard no-bias condition in semiparametric econometrics. When this condition may fail, we introduce a debiased distribution-aligned estimator (DDA) that is suitable when the conditional distribution is learned using deep neural networks. We establish asymptotic theory and introduce a multiplier bootstrap procedure for inference. Simulations support our theory.
\vspace{10pt}

\textbf{Keywords}: distribution alignment; surrogate outcomes; deep neural networks; conditional sampler; multiplier bootstrap; prediction-powered inference; agentic economic modeling; generalized method of moments
\end{abstract}

\newpage

\section{Introduction}\label{sec:intro}

\subsection{Motivation}

The use of proxies for unobserved variables is ubiquitous in empirical research. Classical examples include surrogate biomarkers \citep{Prentice1989SurrogateEndpoints,RobinsRotnitzkyZhao1994MissingRegressors} and measurement error models \citep{chen2005MeasurementErrorModels} with validation data. Large language models (LLMs) are increasingly used to produce proxies for outcomes that are expensive to measure directly. A common workflow pairs a small sample containing true outcome labels with LLM-generated surrogate outcomes available for every observation. Such surrogates can be highly informative, but naively substituting them for the true outcome typically changes the estimand, inducing bias and invalidating inference. At the same time, it is often feasible to collect a small labeled sample containing the true outcome \citep{ji2025PredictionsSurrogates,zhang2025AgenticEconomicModeling,lu2026GAI}.

Across these settings, the central challenge is to combine a large surrogate (``unlabeled'') sample with a smaller labeled sample in a way that is both statistically valid and efficient. In response to the widespread use of AI-generated surrogates, statisticians and machine-learning researchers have developed prediction-powered inference (PPI) and its variants \citep{angelopoulos2023PredictionpoweredInference,angelopoulos2024PPIEfficientPPI,ji2025PredictionsSurrogates}. This line of work developed largely independently of earlier results on measurement error models with validation data, which feature the same two-sample structure  \citep{chen2005MeasurementErrorModels,chen2008SemiparametricEfficiencyGMM}. Furthermore, using a similar motivation, \citet{zhang2025AgenticEconomicModeling} propose agentic economic modeling (AEM). This approach differs substantially from the PPI literature. The objective of this paper is to provide a unified framework for understanding and connecting the methods developed across these communities.

\subsection{Model, Estimators, and Contributions}

In this paper, we study this data combination problem within a parametric \(M\)-estimation framework. The parameter of interest is defined via
\begin{equation}
\theta_0:=\arg\min_{\theta\in\Theta}\mathbb E[\ell_\theta(X^L,Y^L)],
\label{eq:theta0-def}
\end{equation}
where $\ell_\theta(\cdot,\cdot)$ is a loss function, $\Theta\subset\mathbb{R}^p$ is the parameter space, and $(X^L,Y^L)$ is observed in the labeled sample. Let $\widehat Y^L$ denote a surrogate variable generated by a pre-trained model for the labeled data, and let $F_0(y\mid X^L,\widehat Y^L)$ denote the conditional distribution function of $Y^L$ given $(X^L,\widehat Y^L)$. 

In addition to the labeled sample, we also observe an unlabeled sample with $(X^U,\widehat Y^U)$. Throughout we assume the unlabeled and labeled samples are drawn from the same population. We discuss distributional differences between labeled and unlabeled data in Remark~\ref{remark:DistDiff}. In particular, $(X^U,\widehat Y^U)\overset{d}{=}(X^L,\widehat Y^L)$ and the conditional distribution of the (unobserved) $Y^U$ given $(X^U,\widehat Y^U)$ coincides with its labeled version. Under these conditions, $\theta_0$ also admits the following characterization
\begin{equation}
\theta_0=\arg\min_{\theta\in\Theta}\left[\mathbb E\,\ell_\theta(X^L,Y^L)+\mathbb E\left\{\int \ell_{\theta}(X^U,y)\, dF_0(y\mid X^U,\widehat Y^U)\right\}\right].
\label{eq:eq12}
\end{equation}

We construct a \emph{distribution-aligned} (DA) $M$-estimator of $\theta_0$ based on \eqref{eq:eq12}. It incorporates information from the surrogate via the conditional distribution $F_0(\cdot\mid X,\widehat Y)$ and uses the unlabeled sample through the outer expectation in the second term. Because $F_0$ is unknown and must be estimated nonparametrically, \eqref{eq:eq12} defines a semiparametric two-step problem. We estimate $F_0$ from labeled observations and then form a sample analog of the second term using the unlabeled sample, plugging in the estimated conditional distribution. In this way, the conditional distribution of the unobserved $Y^U$ in the unlabeled sample is aligned with that of $Y^L$ in the labeled sample. Intuitively, the conditional distribution captures the full information.

Our DA estimator offers an alternative to prediction-powered inference (PPI) and attains the same efficiency as the current benchmark in this literature \citep{ji2025PredictionsSurrogates}. Consider the formulation of a prototypical PPI estimation derived from the following population-level criterion \citep{zrnic2024CrosspredictionpoweredInference}:
\begin{equation}
\theta_0=\arg\min_{\theta\in\Theta}\left[\mathbb E[\ell_\theta(X^L,Y^L)]+\mathbb E[g_{\theta}(X^U,\widehat Y^U)]-\mathbb E[g_{\theta}(X^L,\widehat Y^L)]\right],
\label{eq:PopPPI}
\end{equation}
which uses unlabeled data via the augmented loss function $g_{\theta}$. The subtraction via the labeled data is to restore consistency and valid inference given the general discrepancy between $Y$ and $\widehat{Y}$. For a particular augmented loss function, PPI does not necessarily outperform the standard M-estimator that makes use of the labeled data alone \citep{angelopoulos2023PredictionpoweredInference}. This issue motivated subsequent work on the search for an efficient augmentation term. See \citet{ji2025PredictionsSurrogates} and references therein, as well as our Section \ref{subsec:PPI} for further details. In contrast to the delicate construction of the efficient augmented term plus the additional bias correction in PPI-type estimators, our DA augments the labeled data with the integrated loss from the unlabeled data. This could be viewed as one particular choice of the augmented loss. The new insight is that the estimation of the nuisance function, i.e., the conditional distribution function, internalizes efficient augmentation. With formal asymptotic expansion, we show that the first-stage estimation error of the conditional distribution propagates into the second-step \(M\)-estimator that produces the same residual score (score function subtracting its conditional expectation) that appears in efficient PPI construction.

Admittedly, the choice of our augmentation via the integrated loss requires either numerical integration with respect to the estimated conditional distribution or making Monte Carlo draws from a fitted conditional sampler \citep{baptista2024ConditionalSamplingMonotoneGANs}. This motivates the use of modern generative modeling tools \citep{Tomczak2024Deep}, including many powerful ones based on deep neural networks (DNN). The efficiency of our DA estimator relies on a no-bias condition \citep{vdVaart1998AsympStat} for the conditional distribution estimator. For highly complex conditional distribution estimators, including DNN based ones that may violate this condition, we also develop a debiased version of our estimator (DDA) that removes the first-order bias. This step makes the source of bias transparent and clarifies how to remove it via debiasing, inspired by the popular DML literature (\citet{ChernozhukovEtAl2018DML}). This also re-frames the debiasing step in PPI-related proposals as a local robustification device \citep{ChernozhukovEtAl2022LocallyRobust} with respect to the unknown conditional distribution, rather than driven solely by the discrepancy between the true outcome and its surrogate.

To sum up, we make several contributions. First, we provide a unified framework that connects semiparametric two-step estimation with prediction-powered inference (PPI) and measurement-error methods with validation data, and that offers a benchmark for comparing with approaches outside the core PPI framework, such as agentic economic modeling (AEM) \citep{zhang2025AgenticEconomicModeling}. Second, we develop asymptotic theory that tracks how first-step estimation error propagates into the second-step \(M\)-estimator. The resulting expansion recovers the efficient augmentation from PPI. The estimated conditional distribution is reusable. Once learned, it can be paired with different loss functions and applied to various estimation tasks for the same data. Third, we show that distributional alignment can be computationally attractive when efficient scores are difficult to derive or to learn for structural econometric models. Finally, we propose a multiplier bootstrap procedure for inference that avoids repeatedly refitting the conditional distribution (or drawing large Monte Carlo samples).

\subsection{Related Literature}\label{subsec:related-lit}

Our paper is most closely related to three branches of the literature: (i) missing-data problems with surrogate measures, (ii) measurement error with validation data, and (iii) prediction-powered inference (PPI).

Classical work on surrogate outcomes and missing data shows that auxiliary variables can improve efficiency through augmented estimating equations \citep{Prentice1989SurrogateEndpoints,RobinsRotnitzkyZhao1994MissingRegressors,WangChen2009EmpiricalLikelihoodMissing}. \citet{RobinsRotnitzkyZhao1994MissingRegressors} develop general theory for semiparametrically efficient estimation and carefully distinguish efficiency for a fixed estimating equation from full semiparametric efficiency (see their Sections 2.6 and 4.2). Our approach is especially close to \citet{WangChen2009EmpiricalLikelihoodMissing}, who estimate a conditional distribution via local constant smoothing within an empirical likelihood framework. We formulate the problem as a generic two-step semiparametric estimator that accommodates a wide range of first-stage estimators. For highly adaptive conditional samplers (e.g., deep-learning-based methods), we additionally develop a debiasing approach that does not rely on the standard no-bias condition required by kernel smoothing estimators.

Our modeling framework is also connected to the validation-sample design in measurement error problems \citep{chen2005MeasurementErrorModels,chen2008SemiparametricEfficiencyGMM}. In that literature, an error-prone proxy $\widehat Y$ for a latent target $Y$ is observed in a large primary sample (our unlabeled sample), along with a smaller validation sample (our labeled sample) in which $Y$ is observed. General two-step GMM and sieve-based estimators are studied by \citet{chen2005MeasurementErrorModels,chen2008SemiparametricEfficiencyGMM}, allowing for different marginal covariate distributions across samples. We show that our DA estimator attains the same efficiency benchmark when the marginal distributions coincide, and we develop the debiased DDA variant to accommodate flexible first-stage conditional distribution estimators as in the contemporary deep learning literature \citep{ChaiFanGu2026NDRE}.

Finally, the PPI literature adapts ideas from \citet{RobinsRotnitzkyZhao1994MissingRegressors} to black-box predictions in modern settings \citep{angelopoulos2023PredictionpoweredInference,angelopoulos2024PPIEfficientPPI,zrnic2024CrosspredictionpoweredInference}. \citet{ji2025PredictionsSurrogates} characterize the optimal conditional score. Their RePPI estimator learns this score directly and uses cross-fitting to accommodate flexible learners. As the PPI literature is growing exponentially, it is impossible to survey all papers. We would like to mention the recent development by \cite{wang2025ELMeetsPPI} on the use of empirical likelihood, \cite{DuanPelger2025Imputation,DuanPelger2026AI} on over-identified moment models with missing covariates under adaptive labeling design, and \cite{Kato2025SSTreatment} for estimating treatment effects with semi-supervised learning. Our DA estimator is a semiparametric plug-in procedure, whereas DDA is its debiased counterpart. This plug-in/debiasing distinction provides a useful benchmark for organizing approaches outside the core PPI framework, and it can be attractive in structural econometric settings \citep{HortacsuJoo2023SE} where deriving or learning efficient scores is computationally difficult.

\subsection{Organization and Notation}\label{subsec:org-notation}

The rest of the paper is organized as follows. Section~\ref{sec:da} introduces the sampling framework and presents our DA estimator, its debiased counterpart, with four illustrative examples. Section~\ref{sec:theory} states regularity conditions and establishes our main asymptotic results. We also compare the no-bias condition underlying DA and the local robustness of DDA therein. Furthermore, we develop a multiplier bootstrap procedure and establish its consistency. Section~\ref{sec:comparisons} connects our approach to PPI and RePPI, AEM, and optimal GMM in measurement error models with validation data. Section~\ref{sec:numerical} presents Monte Carlo results. Section~\ref{sec:conclusion} concludes. The appendices collect the proofs of the main results, further verification of the regularity conditions, technical details on conditional distribution estimation and the connection with optimal GMM. We also provide the detailed computational algorithms for our estimation, its cross-fitted version, and the corresponding bootstrap procedure.

We write $Z=(Y,X,\widehat Y)$ for a labeled observation and $V=(X,\widehat Y)$ for an unlabeled one. Expectations under the labeled and unlabeled populations are denoted by $P_0$ and $P_V$ (when they coincide, we suppress the distinction). The empirical measures based on the labeled and unlabeled samples are denoted by $\mathbb P_n$ and $\mathbb Q_N$. For a measurable function $f$, we write $\|f\|_{P,2}=(P f^2)^{1/2}$. We use $\op(1)$ and $\Op(1)$ for convergence in probability and stochastic boundedness, respectively. We write $\rightsquigarrow$ for the weak convergence \citep{vdVaartWellner2023EP}.

\section{Distributional Alignment in \(M\)-Estimation}\label{sec:da}

We observe a labeled sample $\{(Y_i,X_i,\widehat Y_i)\}_{i=1}^n$ and an independent unlabeled sample $\{(X_j,\widehat Y_j)\}_{j=n+1}^{n+N}$. We drop the superscript L or U on the random variables in the estimation part, as the subscripts over different samples can differentiate them. Here $X_i\in\mathcal X\subset\mathbb R^{d_X}$, $Y_i\in\mathcal Y\subset\mathbb R^{d_Y}$, and $\widehat Y_i$ is a surrogate outcome produced by a pre-trained model.\footnote{The setup extends directly to $K$ pre-trained models producing $\widehat{Y}_{\ell,k}$ for $\ell=1,\ldots,n+N$ and $k=1,\ldots,K$. Collect these into the vector surrogate $\widehat{Y}_{\ell}=\bigl(\widehat{Y}_{\ell,1},\ldots,\widehat{Y}_{\ell,K}\bigr)$ and condition on this vector throughout.} We view the surrogate generator as trained on external data, so that (conditional on its training data) the observations remain independent and identically distributed (\textit{i.i.d.})).

The target parameter $\theta_0$ is defined in \eqref{eq:theta0-def} for a given loss $\ell_\theta$ and is assumed to be point identified from the distribution of $(X,Y)$. Using labeled data alone, $\theta_0$ can be estimated by the conventional \(M\)-estimator:
\begin{equation}
\widehat{\theta}_M
:=\arg\min_{\theta\in\Theta}\frac{1}{n}\sum_{i=1}^n \ell_{\theta}(X_i, Y_i).
\end{equation}

In order to utilize the unlabeled sample together with the surrogate outcomes $(X_j,\widehat{Y}_j)_{j=n+1}^{n+N}$, a naive approach treats $\widehat{Y}_j$ as if it were the true outcome and estimates $\theta_0$ via
\begin{equation}
\widehat{\theta}_{\mathrm{Naive}}
:=\arg\min_{\theta\in\Theta}\left(\frac{1}{n}\sum_{i=1}^n\ell_{\theta}(X_i, Y_i)+\frac{1}{N}\sum_{j=n+1}^{n+N}\ell_{\theta}(X_j, \widehat{Y}_j)\right).
\label{Naive}
\end{equation}
However, $\widehat{\theta}_{\mathrm{Naive}}$ is generally biased \citep{angelopoulos2023PredictionpoweredInference,zrnic2024CrosspredictionpoweredInference}. This motivates recent work on debiasing $\widehat{\theta}_{\mathrm{Naive}}$ and on improving the efficiency of $\widehat{\theta}_M$. For M-estimation, PPI and its extensions have been studied extensively. See subsection \ref{subsec:PPI} for a brief discussion.

Following the PPI literature, we impose the following sampling assumption throughout. 

\begin{assumption}[Sampling scheme]\label{Assump:Sampling}
The triples $\{(X_i,Y_i,\widehat Y_i)\}_{i=1}^{n+N}$ are i.i.d.\ with joint law $P_0$, where $\{Y_i\}_{i=n+1}^{n+N}$ are not observed in the sample. We assume the missing completely at random (MCAR). Moreover, $r_n:=n/N\to r\in[0,\infty)$ as $n,N\to\infty$. 
\end{assumption}

\subsection{A New Estimator via Distributional Alignment}\label{subsec:new-estimator}
Let
\[
 F_0(y\mid x,\widehat y)
 =P(Y^L\le y\mid X^L=x,\widehat Y^L=\widehat y)
\]
denote the true conditional distribution of $Y^L$ given $(X^L, \widehat Y^L)$. Under our Assumption \ref{Assump:Sampling}, we have the compatibility of two conditional distributions as $P(Y^L\le y\mid X^L=x,\widehat Y^L=\widehat y)=P(Y^U\le y\mid X^U=x,\widehat Y^U=\widehat y)$. For any distribution $F(\cdot\mid x,\widehat y)$, define the conditional expected loss
\begin{align}
\mu_\theta(x,\widehat y;F)
 :=\int\ell_\theta(x,y)\,dF(y\mid x,\widehat y).
 \label{eq:mu-def}
\end{align}
Under \cref{Assump:Sampling}, Equation~\eqref{eq:eq12} holds and can be written as
\begin{equation}
 \theta_0=\arg\min_{\theta\in\Theta}
 \mathbb E
       [\ell_\theta(X^L,Y^L)]+\mathbb E
       [\mu_\theta(X^U,\widehat Y^U;F_0)].
 \label{eq:eq12-mu}
\end{equation}
It suggests a two-step semiparametric estimation in which the nuisance function is the conditional distribution $F_0$. This characterization of $\theta_0$ offers an alternative route distinct from the existing PPI-type adjustment.

Specifically, define the population criterion
\begin{equation}\label{eqn:DACriterion}
    Q_{\mathrm{DA}}(\theta):=\mathbb{E}\bigl[\ell_{\theta}(X^L,Y^L)\bigr]+\mathbb{E}\bigl[\mu_{\theta}(X^U,\widehat Y^U;F_0)\bigr].
\end{equation}
We estimate the integrated loss $\mu_{\theta}(X^U,\widehat Y^U;F_0)$ by a plug-in approach. First, we estimate $F_0$ from the labeled sample and obtain $\widehat F$. We then form the plug-in estimator $\mu_{\theta}(X^U,\widehat Y^U;\widehat F)$ and estimate $\theta_0$ by
\[
\widehat\theta_{\mathrm{DA}}\in\arg\min_{\theta\in\Theta}\widehat Q_{\mathrm{DA}}(\theta),
\]
where
\begin{align}
\widehat Q_{\mathrm{DA}}(\theta)
 :=\frac1{n+N}\left\{
       \sum_{i=1}^n\ell_\theta(X_i,Y_i)
       +\sum_{j=n+1}^{n+N}\mu_\theta(X_j,\widehat Y_j;\widehat F)
\right\}.
 \label{eq:DA-estimator}
\end{align}

The DA estimator is therefore a semiparametric two-step procedure whose first-step nuisance is the conditional distribution of the true outcome given covariates and surrogates. We use the term \emph{distributional alignment} because the conditional law estimated from labeled data is used to impute outcomes in the unlabeled sample.

For a general loss function, evaluating $\mu_\theta(X_j,\widehat Y_j;\widehat F)$ may require simulation. For each unlabeled pair $(X_j,\widehat Y_j)$, draw
$Y_j^{(s)}\sim\widehat F(\cdot\mid X_j,\widehat Y_j)$, $s=1,\ldots,S$, independently across $j$ and $s$ conditional on the data\footnote{\cite{DuanPelger2025Imputation} examine the settings where each missing covariate is replaced by \textit{one} imputation per units. Besides the debiasing methodology that is related to PPI, their focus is on adaptively reweighting the imputed values based on the imputation quality.}. This yields the simulated DA (SDA) criterion
\begin{equation}
\widehat\theta_{\mathrm{DA},S}\in\arg\min_{\theta\in\Theta}
 \frac1{n+N}\left\{
       \sum_{i=1}^n\ell_\theta(X_i,Y_i)
       +\frac1S\sum_{j=n+1}^{n+N}\sum_{s=1}^S
                       \ell_\theta(X_j,Y_j^{(s)})\right\}.
 \label{eq:SDA-estimator}
\end{equation}
Repeated sampling from an estimated conditional distribution can pose a computational challenge for classical nonparametric estimators. Deep generative models, particularly conditional samplers, are well suited to this task \citep{Tomczak2024Deep}. Deep neural networks accommodate flexible distribution functions, while carefully designed architectures permit efficient generation of draws. We analyze the integrated DA estimator (equivalently, the limit of $\widehat\theta_{\mathrm{DA},S}$ as $S\to\infty$) defined in \eqref{eq:DA-estimator}. For fixed $S$, simulation noise can be handled using standard multiple-imputation adjustments\footnote{The transition from a finite $S$ to the theoretical infinite one is well understood; see the asymptotic behavior of the $A_n$ term in \citet{WangChen2009EmpiricalLikelihoodMissing} and their Proposition A.1.}.

Alternatively, one can estimate the integrated loss in \eqref{eq:mu-def} directly by regressing $\ell_\theta(X_i,Y_i)$ on $(X_i,\widehat Y_i)$. Let $\widehat\mu_\theta(x,\widehat y)$ denote the fitted conditional mean. This regression-based implementation, which we call regression distributional alignment (RDA), proceeds as follows:
\begin{equation}
 \begin{split}
 \widehat\theta_{\mathrm{RDA}}
   &\in\arg\min_{\theta\in\Theta}\widehat Q_{\mathrm{RDA}}(\theta),\\
 \widehat Q_{\mathrm{RDA}}(\theta)
   &:=\frac1{n+N}\left\{
       \sum_{i=1}^n\ell_\theta(X_i,Y_i)
       +\sum_{j=n+1}^{n+N}
                    \widehat\mu_\theta(X_j,\widehat Y_j)\right\}.
 \end{split}
 \label{eq:RDA-estimator}
\end{equation}
This avoids repeated simulation draws, but it may require refitting a regression for each candidate $\theta$, resembling the iterated/profiled MLE \citep{ichimura1993single,ichimura2010semi} in semiparametric estimation problems. For the generalized linear models (GLMs) considered below, this is not necessary: one only needs to estimate the conditional mean of $Y$ given $(X,\widehat Y)$. To avoid repetition, we focus on $\widehat{\theta}_{\mathrm{DA}}$ in our paper.

For subsequent analysis, we set
\[
 a_n=\frac n{n+N}=\frac{r_n}{1+r_n},\qquad
 b_n=\frac N{n+N}=\frac1{1+r_n}.
\]
We also adopt the following empirical process notations:
\[
 \begin{split}
 \mathbb P_n f(X,Y,\widehat Y)
   :=\frac1n\sum_{i=1}^n f(X_i,Y_i,\widehat Y_i),~~~
 \mathbb Q_N g(X,\widehat Y)
   :=\frac1N\sum_{j=n+1}^{n+N}g(X_j,\widehat Y_j).
 \end{split}
\]
In particular, we have 
\begin{equation}
 \widehat Q_{\mathrm{DA}}(\theta)
 =a_n\mathbb P_n\ell_\theta(X,Y)
      +b_n\mathbb Q_N\mu_\theta(X,\widehat Y;\widehat F),
 \label{eq:DA-weighted-criterion}
\end{equation}
as the sample criterion function in our DA estimation.

\subsection{Debiased Distributional Alignment Estimator}\label{subsec:dda}

As shown in Section \ref{sec:theory}, our DA estimator is root-$n$ consistent and asymptotically normal under standard regularity conditions for two-step semiparametric estimation. As such, it requires an asymptotic linear representation of certain functional of the estimated conditional distribution $\widehat{F}$. For highly complicated conditional distribution estimator, this condition may not be straightforward to verify.

Motivated by the double/debiased machine learning (DML) literature \citep{ChernozhukovEtAl2018DML}, we consider an explicit correction for the first-order effect of estimating the conditional distribution. We refer to the resulting estimator as \emph{debiased distributional alignment} (DDA):
\begin{equation}
    \thetabc
    \in\arg\min_{\theta\in\Theta}
    \widehat Q_{\mathrm{DDA}}(\theta),
    \label{eq:DDA-estimator}
\end{equation}
where 
\begin{align}
    \widehat Q_{\mathrm{DDA}}(\theta)
 &=\mathbb P_n\ell_\theta(X,Y)
   +b_n\{\mathbb Q_N\mu_\theta(X,\widehat Y;\widehat F)
             -\mathbb P_n\mu_\theta(X,\widehat Y;\widehat F)\}
       \label{eq:DDA-criterion}\\
 &=\widehat Q_{\mathrm{DA}}(\theta)
   +b_n\mathbb P_n\{\ell_\theta(X,Y)
                        -\mu_\theta(X,\widehat Y;\widehat F)\}.
       \label{eq:DDA-alternative}
\end{align}
The second expression shows that DDA adds the labeled loss residual to the DA criterion. Equivalently, the first expression centers the augmentation
$b_n\mu_\theta(X,\widehat Y;\widehat F)$ across the two samples.

Compared with the original DA criterion \eqref{eq:DA-estimator}, the alternative expression in \eqref{eq:DDA-alternative} shows that DDA uses the centered loss $\Pn\{\ell_\theta-\mu_\theta(\,\cdot\,;\widehat F)\}$ as an explicit debiasing.\footnote{As in DML, cross-fitting can be used when $\widehat F$ is obtained by a highly adaptive learner. The trade-off is that sample splitting reduces the effective labeled sample size available to fit the conditional distribution when $n$ is small. In our main theoretical results, we focus on the no-splitting case.} This construction also offers a complementary perspective on the PPI literature. PPI methods begin with the naive estimator \eqref{Naive}, correct its bias due to the discrepancy between $Y$ and $\widehat Y$, and then search for an efficient augmentation. In contrast, DA begins with an efficient plug-in estimator based on an estimated conditional distribution, and DDA then removes the first-order bias induced by estimating that nuisance function, thereby avoiding the no-bias condition.

Despite that DDA generally requires weaker conditions in the asymptotics, it remains insightful to examine the DA estimator as a starting point. The reason is that it serves as the efficient plug-in type approach, which provides the benchmark to compare other estimators, rather than just the PPI class. The other nice feature is that DA naturally preserves the convexity of $\widehat{Q}_{DA}$, when the sample loss function $\ell_{\theta}$ is convex. Since $\mu_{\theta}(\cdot;F)$ is the conditional expectation of $\ell_{\theta}$, it is also convex in $\theta$. In comparison, the DDA objective \eqref{eq:DDA-criterion} need not be convex\footnote{Note that the linearized implementation of \citet{ji2025PredictionsSurrogates} replaces the imputed loss by an affine function of $\theta$, thereby recovering a globally convex optimization problem.}.

We describe two simple ways to handle the non-convexity of our DDA. A first route is to use the DA estimator as an initial value and then perform one or several updates, e.g., damped Newton or quasi-Newton updates when the loss is differentiable. In practice, one can also use the labeled-only estimator as a cheap second start and retain the solution with the smaller DDA objective. A second route exploits the additional structure whenever the original loss is convex. This convex loss structure also gives the difference-of-convex decomposition $\widehat Q_{\mathrm{DDA}}(\theta)=\mathcal A_n(\theta)-\mathcal B_n(\theta)$, where
\[
 \mathcal A_n(\theta)=\mathbb P_n\ell_\theta(X,Y)
          +b_n\mathbb Q_N\mu_\theta(X,\widehat Y;\widehat F),~~~
 \mathcal B_n(\theta)=b_n\mathbb P_n\mu_\theta(X,\widehat Y;\widehat F).
\]
Both components are convex, permitting efficient difference-of-convex algorithms \citep{PhamLe2014DCA}. We refer readers to Section 4.4.4 of \cite{CuiPang2021Optimization} for a wealth of fundamental results in this area. 
\begin{remark}
    We connect to the classical missing data literature \citep{RobinsRotnitzkyZhao1994MissingRegressors} and compare two proposals that we have. Similar formulations for the RePPI and GAI approaches are given by \cite{ji2025PredictionsSurrogates} and \cite{lu2026GAI} respectively. Introduce the dummy variable $D_k=1$ for $k\le n$ and $D_k=0$ otherwise. As it becomes clear, both approaches agree with our DDA in the absence of sample splitting or cross fitting tricks. The sample criterion of our DA estimator is
    \begin{align}
         \widehat Q_{\mathrm{DA}}(\theta)
 =\frac1{n+N}\sum_{k=1}^{n+N}
 \{D_k\ell_\theta(X_k,Y_k)
       +(1-D_k)\mu_\theta(X_k,\widehat Y_k;\widehat F)\}.
 \label{eqn:MissingDataDA}
    \end{align}
Note that the imputation only applies to the unlabeled data for which we makes use of the estimated conditional distribution, cf. \eqref{eqn:MissingDataDA} with the estimating equation (3) in \cite{WangChen2009EmpiricalLikelihoodMissing}. 

The propensity score is constant in our setup, i.e., we assume MCAR. One can simply estimate it by $ \widehat\pi_n(x,\widehat y)=a_n$. Thereafter, the sample criterion function of our DDA estimator can be expressed as
    \begin{align}
       \widehat Q_{\mathrm{DDA}}(\theta)
 =\frac1{n+N}\sum_{k=1}^{n+N}
 \left[\mu_\theta(X_k,\widehat Y_k;\widehat F)
       +\frac{D_k}{a_n}\{
           \ell_\theta(X_k,Y_k)
                   -\mu_\theta(X_k,\widehat Y_k;\widehat F)\}\right].
 \label{eqn:MissingDataDDA}
    \end{align}
    In the last equality, we can see that the imputation applies to all observations, whereas the second term corrects the estimation effect of the conditional distribution. One can check that the orthogonal score in equation (7) of \cite{lu2026GAI} agrees with \eqref{eqn:MissingDataDDA}, except for a varying propensity score function that depends on covariates. 
\end{remark}

\subsection{Illustrative Examples}\label{sec:examples} 

It is instructive to illustrate our methodology using concrete models. The first three examples are widely used in the PPI literature \citep{angelopoulos2023PredictionpoweredInference,angelopoulos2024PPIEfficientPPI,zrnic2024CrosspredictionpoweredInference}. In these settings, the integrated loss reduces to a conditional mean, so the implementation amounts to a (nonparametric) regression rather than extensive simulation. We denote the conditional expectation by
\[
 \nu_0(x,\widehat y):=\E\bigl[Y\mid X=x,\widehat Y=\widehat y\bigr],
 \qquad
 \widehat\nu(x,\widehat y):=\int y\, d\widehat F\bigl(y\mid X=x,\widehat Y=\widehat y\bigr).
\]
\begin{example}[Mean Estimation]
\label{ex:mean-calibration}
Consider the standard mean estimation of the outcome. The target is $\theta_0=\E[Y]$ and the loss function is $\ell_\theta(X,Y)=\frac12(Y-\theta)^2$. 
It is straightforward to derive that our DA estimator is
\begin{equation}
 \widehat\theta_{\rm DA}
 =a_n\Pn Y+b_n\QN\widehat\nu.
 \label{eq:mean-DA}
\end{equation}
\end{example}

\begin{example}[Linear Regression and GLMs]
\label{ex:glm-calibration}
Consider the linear regression with the following squared distance and its score as:
\[
 \ell_\theta(X,Y)=\frac12(Y-X^\top\theta)^2,
 \qquad
 s_\theta(X,Y):=\partial_{\theta} \ell_\theta(X,Y)=X(X^\top\theta-Y).
\]
By straightforward algebra, the DA estimator can be written as
\begin{align}
 \widehat\theta_{\rm DA}
 &=\{\widehat\Sigma_{XX}^{\rm DA}\}^{-1}
    \widehat\Sigma_{XY}^{\rm DA},
 \label{eq:OLS-DA}\\
 \widehat\Sigma_{XX}^{\rm DA}
 &:=a_n\Pn(XX^\top)+b_n\QN(XX^\top),\\
 \widehat\Sigma_{XY}^{\rm DA}
 &:=a_n\Pn(XY)+b_n\QN\{X\widehat\nu\}.
\end{align}
The same structure persists, whenever the score function corresponding to the loss function is affine in the outcome which covers the canonical generalized linear model. Recall for the (negative) quasi-likelihood criterion:
\begin{equation}
 \ell_\theta(X,Y)=A(X^\top\theta)-Y X^\top\theta,
 \label{eq:canonical-GLM-loss}
\end{equation}
with a given convex link function $A(\cdot)$. In this case, we have
\begin{align}
 \mu_\theta(V;\widehat F)=A(X^\top\theta)-\widehat\nu X^\top\theta,\label{eq:GLM-integrated-loss}
\end{align} 
Note that our DA preserves the convex minimization structure.
\end{example}

The simplification for the first two examples does not carry over to more complicated models. For either DA or DDA, the distributional alignment step for the quantile regression differs operationally from the PPI-type approach in \cite{zrnic2024CrosspredictionpoweredInference}. 
\begin{example}[Quantile Regression]
Consider the linear quantile regression model for a given quantile level $\tau\in(0,1)$:
\begin{align*}
    Y_i = X_i^{\top} \theta_0 + \epsilon_i, \quad F_{\epsilon|x}^{-1}(\tau|X_i) = 0.
\end{align*}
The quantile regression coefficient is estimated by minimizing the check function $\ell_{\theta}(X,Y) = \rho_\tau(Y - X^{\top} \theta)$ in which $ \rho_{\tau}(u) = u (\tau - 1(u \le 0))$ \citep{KoenkerBassett1978Quantile}. Our simulated DA estimator naturally preserves the convexity of the minimization problem by solving
\begin{equation*}
    \min_{\theta}\frac{1}{n+N}\left\{\sum_{i=1}^n \rho_\tau(Y_i - X_i^{\top} \theta)+ \frac{1}{S}\sum_{j=n+1}^{n+N} \sum_{s=1}^S\rho_\tau(Y^{(s)}_j - X_i^{\top} \theta)\right\},
\end{equation*}
where for each unlabeled pair $(X_j,\widehat Y_j)$, we make independent draws from the estimated conditional distribution, i.e.,
$Y_j^{(s)}\sim\widehat F(\cdot\mid X_j,\widehat Y_j)$, $s=1,\ldots,S$ and $j=n+1,\cdots,n+N$. 
\end{example}

In the final example prevalent in structural econometrics, evaluating the sample score function requires differentiating the solution to a fixed-point problem, which makes the proposal in \cite{ji2025PredictionsSurrogates} computationally prohibitive. Our DA approach avoids setting this derivative (score) as a separate target. Instead, we first estimate $\widehat F$ and then plug simulated outcomes from the estimated conditional distribution into the original criterion function, while the Bellman equation and the associated structural computation remain unchanged within the existing fixed-point loop.

\begin{example}[Dynamic Discrete Choice]\label{ex:structural-complicated-likelihood}
Consider the single-agent dynamic discrete choice model in \cite{Rust1987BusEngines}. The agent maximizes the discounted sum of the utilities $u_{\theta}(a,s)$, which depends on the parameter $\theta$, the action $a$ and the state variable $s$. Following \cite{Rust1987BusEngines}, there is an unobserved component $\epsilon$ that follows type I extreme value distribution. The agent chooses $A_t\in\mathcal A$ in state $S_t$, with a value function $V_{\theta}(\cdot)$ satisfying the Bellman equation:
	\begin{equation}
		V_{\theta}(s)
		=
		\mathbb{E}_{\varepsilon}
		\left[
		\max_{a\in\mathcal A}
		\left\{
		u_{\theta}(a,s)+\varepsilon_a
		+
		\delta
		\mathbb{E}_{\theta}
		[V_{\theta}(S_{t+1})\mid S_t=s,A_t=a]
		\right\}
		\right],
		\label{eq:ddc-bellman}
	\end{equation}
with a known discount factor $\delta$. Under the set of identification conditions in \cite{Rust1987BusEngines}, we can express the implied conditional choice probability $P_{\theta}(a\mid s)$ as a soft-max function depending on the value function $V_{\theta}$\footnote{ See Section 2.3 of \cite{HortacsuJoo2023SE} for the comprehensive exposition of this model.}. For illustration, we assume the access to some unlabeled data where the discrete choices are missing from period $t=2,\cdots,T$, so that we take $(A_{t})_{t=2}^T$ as the missing outcome $Y$ and make use of AI surrogate instead. If one starts with the maximum likelihood estimator of \cite{Rust1987BusEngines}, a recalibration method as in \cite{ji2025PredictionsSurrogates} requires fitting a regression of the sample score function which in turn depends on the score from the conditional choice probabilities:
	\begin{equation}
		\partial_{\theta}
		\log P_{\theta}(A_{it}\mid S_{it})
		\bigg|_{\theta=\widehat\theta_I}, ~~\text{for}~~t=2,\cdots,T.
		\label{eq:ddc-score}
	\end{equation}
Because $P_{\theta}$ is itself determined by the value function in Equation \eqref{eq:ddc-bellman}, constructing \eqref{eq:ddc-score} can be practically challenging. In comparison, our approach would fit the conditional distribution of the outcome on the covariates and surrogates, and we keep generating draws of missing outcomes for the unlabeled data. 
\end{example}

\section{Main Theoretical Results}\label{sec:theory}
In this section, we establish the asymptotic theory of our estimators. We introduce some additional notations that facilitate our presentation. We use the shorthand $V=(X,\widehat Y)$ to collect covariates and surrogates. The labeled sample consists of $Z_i=(Y_i,V_i)$ for $i=1,\ldots,n$, and the (independent) unlabeled sample consists of $V_j$ for $j=n+1,\ldots,n+N$. Referring to empirical averages as $\Pn f=\frac1n\sum_{i=1}^n f(Z_i)$ and $\QN g=\frac1N\sum_{j=n+1}^{n+N}g(V_j)$, we write the corresponding (centered and normalized) empirical processes $\G_n f:=\frac1{\sqrt n}\sum_{i=1}^n\{f(Z_i)-P_0f\}$ and $\G_N g:=\frac1{\sqrt N}\sum_{j=n+1}^{n+N}\{g(V_j)-P_Vg\}$. When $g$ is a function of $V$ alone, $\Pn g=n^{-1}\sum_{i=1}^n g(V_i)$ and $P_0g=P_Vg$. 

As in the classical \(M\)-estimation problem \citep{vdVaart1998AsympStat}, the score function plays an important role. In the smooth case, define the score function and its conditional expectation as follows:  
\begin{align}
    s_\theta(x,y)
	&:=\partial_\theta\ell_\theta(x,y), \label{eq:score-def}\\
	\varsigma_\theta(v;F)
	&:=\int s_\theta(x,y)\,dF(y\mid v). \label{eq:m-def}
\end{align}
Our general theory allows $\ell_\theta$ to be non-smooth in $\theta$. In this case, the above functions are defined via the stochastic differentiability in \eqref{eq:SE-labeled}, similar to \cite{PakesPollard1989Simulation}. Specifically for convex non-smooth losses, $s_\theta$ denotes a measurable subgradient \citep{Pollard1991LAD}. These two functions at the truth are simply denoted by $s_0(Z):=s_{\theta_0}(X,Y)$ and $\varsigma_0(V):=\E[s_0(Z)\mid V]$.

Recall that for the parameter $\theta\in \Theta\subset \mathbb R^p$, we denote the corresponding Euclidean norm by $\Vert \cdot\Vert$. For a vector-valued function $g$, set $\|g\|_{P,q}=\left(\int\|g\|^q\,dP\right)^{1/q}$ for $q=1,2$ given the probability measure $P$. For a matrix
$A\in\mathbb R^{p\times p}$, we use the following operator norm
\begin{equation}
 \|A\|_{\mathrm{op}}
 =\sup_{u\in\mathbb R^p,\|u\|=1}\|Au\|
 =\lambda_{\max}(A^\top A)^{1/2}.
\label{eq:bootstrap-matrix-norm}
\end{equation}

\subsection{Regularity Conditions}
We list our regularity conditions, followed by some heuristic discussions.
\begin{assumption}[Identification]\label{Assump:ID}
The true parameter \(\theta_0\) is an interior point of the compact parameter space $\Theta$. The population-level loss function \(Q(\theta):=\E[\ell_\theta(X,Y)]\) admits a unique and well-separated minimizer at \(\theta_0\), i.e., for every \(\epsilon>0\),
$ \inf_{\theta:\|\theta-\theta_0\|\geq\epsilon}\{Q(\theta)-Q(\theta_0)\}>0.$
\end{assumption}
\begin{assumption}[Glivenko-Cantelli]\label{Assump:GC}
	We assume that the functional class $\mathcal{L}:=\{\ell_{\theta}(x,y):\theta\in\Theta \}$ is $P_0$-Glivenko-Cantelli, with a $L_2$-integrable envelope function \(L(Z)\) satisfying \(\E \left(L(Z)^2\right)<\infty\).    
\end{assumption}

\begin{assumption}[First-step Consistency]
	\label{Assump:first-step-L2}
The estimated conditional distribution $\widehat F$ satisfies
\begin{equation}
    P_V\left[
        \sup_{\theta\in\Theta}
        \left|
            \mu_\theta(\cdot;\widehat F)
            -\mu_\theta(\cdot;F_0)
        \right|
    \right]=\op(1).
\end{equation}
Furthermore, let
	\begin{equation}
		\Delta_n(v;\widehat F):=\varsigma_{\theta_0}(v;\widehat F)-\varsigma_0(v).
		\label{eq:Delta-score}
	\end{equation}
We assume that it satisfies
	\begin{equation}
		\|\Delta_n(\cdot;\widehat F)\|_{P_V,2}=\op(1).
		\label{eq:L2-first-step}
	\end{equation}
\end{assumption}

\begin{assumption}[Local Smoothness and Curvature]\label{Assump:Smooth}
For almost every \((x,y)\), the map \(\theta\mapsto\ell_\theta(x,y)\) is convex. The population criterion admits the following quadratic expansion
\begin{equation}
 Q(\theta_0+t)-Q(\theta_0)
 =\frac12 t^\top H t+o(\|t\|^2),\qquad t\to0,
 \label{eq:population-quadratic}
\end{equation}
in a neighborhood of \(\theta_0\), where \(H\) is symmetric positive definite, with its minimum eigenvalue satisfying $\lambda_{min}(H)>0$.
\end{assumption}

\begin{assumption}[Stochastic Differentiability]\label{Assump:SE}
For \(\theta_{n,h}:=\theta_0+h/\sqrt n\) and every fixed \(M<\infty\), there exist some functions $s_0(\cdot,\cdot)$ and $\varsigma_0(v;F_0)=\int s_0(x,y)\,dF_0(y\mid v)$ such that
\begin{align}
&\sup_{\|h\|\le M}
\left|\G_n\!\left[\sqrt n\{\ell_{\theta_{n,h}}-\ell_{\theta_0}\}-h^\top s_0\right]\right|=\op(1),
\label{eq:SE-labeled}\\
&\sup_{\|h\|\le M}
\left|\G_N\!\left[\sqrt n\{\mu_{\theta_{n,h}}(\cdot;F_0)-\mu_{\theta_0}(\cdot;F_0)\}-h^\top \varsigma_0\right]\right|=\op(1),
\label{eq:SE-unlabeled}\\
&\sup_{\|h\|\le M}
\left|\G_n\!\left[\sqrt n\{\mu_{\theta_{n,h}}(\cdot;F_0)-\mu_{\theta_0}(\cdot;F_0)\}-h^\top \varsigma_0\right]\right|=\op(1).
\label{eq:SE-labeled-integrated}
\end{align}
\end{assumption}

\begin{assumption}[Local Linearization]\label{Assump:PlugLocal}
For \(\theta_{n,h}=\theta_0+h/\sqrt n\), define
\begin{align}
D_{n,h}(v;\widehat{F}):={}&\{\mu_{\theta_{n,h}}(v;\widehat F)-\mu_{\theta_0}(v;\widehat F)\}
-\{\mu_{\theta_{n,h}}(v;F_0)-\mu_{\theta_0}(v;F_0)\}.
\label{eq:Dnh-def}
\end{align}
For every fixed \(M<\infty\), we assume that
\begin{equation}
 \sup_{\|h\|\le M}
 \left|n\,\QN D_{n,h}(\cdot;\widehat{F})-h^\top\sqrt n\,\QN\Delta_n(\cdot;\widehat{F})\right|=\op(1).
 \label{eq:local-first-step-increment}
\end{equation}
\end{assumption}

\begin{assumption}[Asymptotic Linear Representation]\label{ass:first-step-AL}
We assume the asymptotic linear representation holds
\begin{equation}
 \sqrt n\,P_V\Delta_n(\cdot;\widehat{F})
 =\G_n\{s_0-\varsigma_0\}+\op(1).
 \label{eq:first-step-AL}
\end{equation}
\end{assumption}

\begin{assumption}[Moments]\label{Assump:Moments}
	For some \(\delta>0\),
$	\E\|s_0(Z)\|^{2+\delta}<\infty$ and $	\E\|\varsigma_0(V)\|^{2+\delta}<\infty.$
\end{assumption}

\begin{assumption}[Local Robustness]\label{Assump:DDA-local}
For every fixed \(M<\infty\), we assume that
\begin{equation}
 \sup_{\|h\|\le M}
 n\left|(\QN-\Pn)D_{n,h}(\cdot;\widehat{F})\right|=\op(1).
 \label{eq:DDA-local-orthogonality}
\end{equation}
\end{assumption}
Assumptions \ref{Assump:ID}, \ref{Assump:GC} to \ref{Assump:first-step-L2} are generic, concerning the identifiability of the model parameter, the uniform law of large number of the sample loss function and the asymptotics of the first stage nonparametric estimation, cf. \cite{ChenLintonVanKeilegom2003Nonsmooth,ichimura2010semi,NanWellner2013SemiparametricZ}. We keep the metric that quantifies the first stage estimation as a high level condition in Assumption \ref{Assump:first-step-L2}. As discussed in Appendix C, this can be verified via the integrated $L_2$-norm \citep{ChaiFanGu2026NDRE}, depending on primitive conditions on the score function. Let $\epsilon_n$ denote the rate of convergence of $\widehat{F}$. It seems that we are not imposing the usual rate condition as $\epsilon_n=o(n^{-1/4})$ \citep{Newey1994AsymptoticVariance} upfront, because the conditional density enters linearly into the criterion function, so there is no explicit second-order term. That said, when we check Assumption \ref{ass:first-step-AL} or \ref{Assump:DDA-local} through stochastic equicontinuity, we still need this type of rate restriction for showing the Donsker property for properly localized empirical processes.

As the sample loss function need not be differentiable in a neighborhood of \(\theta_0\), Assumption \ref{Assump:Smooth} only imposes the local quadratic expansion at the population level \citep{PakesPollard1989Simulation}. Regarding the stochastic differentiability in Assumption \ref{Assump:SE}, note that we impose it with respect to the parametric part, at the true conditional distribution function. Therein, the third display is only needed for DDA. This is standard and can be checked following \cite{PakesPollard1989Simulation,Pollard1991LAD,Sherman1993MRC} for smooth or non-smooth criterion functions. Assumption \ref{Assump:PlugLocal} maintains the stochastic differentiability with the estimated conditional distribution. This can be checked by assuming the estimated conditional distribution function belongs to some Donsker class via standard empirical process theory. The Lyapunov type moment restrictions on the score function and its conditional expectation in Assumption \ref{Assump:Moments} are needed for establishing the central limit theorem.

The non-trivial assumption is Assumption \ref{ass:first-step-AL} for our DA estimator, which states the asymptotic representation of the linear functional of the conditional distribution. Compared to the generic results in \cite{ChenLintonVanKeilegom2003Nonsmooth} or \cite{NanWellner2013SemiparametricZ} on two-stage semiparametric estimation, here we assume the influence function is of the specific form in Assumption \ref{ass:first-step-AL}, i.e., it equals to the residual score function $s_0(Z)-\varsigma_0(V)$. For classical nonparametric estimators, such as kernel or sieve based, this can be shown directly. We refer to \cite{WangChen2009EmpiricalLikelihoodMissing} for the local constant kernel smoothing estimator and \cite{chen2005MeasurementErrorModels} for the linear sieve estimator. We also provide a concrete verification for the local linear estimator used in our Monte Carlo simulation. In addition, we present the formal influence function calculation following \cite{Newey1994AsymptoticVariance} to ensure that this is the form for generic nonparametric first stage estimation. 

The influence function argument alone is not sufficient for justifying the linear representation for any first-stage estimator $\widehat{F}$. Next, we carry out some additional expansion to show that the crux for Assumption \ref{ass:first-step-AL} to hold is the no-bias condition \citep{vdVaart1998AsympStat}. As noted by \cite{ChernozhukovEtAl2022LocallyRobust}, this is a non-trivial restriction and often requires substantial work to check for complex nonparametric estimators. We show that for the GLM, there is a very simple strategy to construct the nonparametric estimator so that this no-bias condition is enforced. Alternatively, one can also use the targeted MLE of \cite{vdLaanRose2011Targeted}.

The above remark sheds light on the robustness of our DDA, which does not require the asymptotic linear representation in Assumption \ref{ass:first-step-AL}. Instead, it relies on the weaker condition in Assumption \ref{Assump:DDA-local}, which we refer to as \emph{local robustness} (cf. \citealt{ChernozhukovEtAl2022LocallyRobust}). It requires the negligibility of certain difference involving the local oscillation $D_{n,h}(v;\widehat{F})$ between two sample averages over the labeled and unlabeled data, which echos the PPI construction. With sample splitting \citep{vdVaart1998AsympStat} or cross-fitting \citep{ChernozhukovEtAl2018DML,ji2025PredictionsSurrogates}, this condition can be shown under standard consistency of $\widehat{F}$. When the full sample is used for first-stage estimation, one route is to assume that the estimated conditional distributions lie in a Donsker class. Even for deep neural network (DNN) estimators, Donsker-type conditions can be verified under structural restrictions on the underlying function class. Such structure is also used to establish DNN convergence rates and represents a significant recent advance \citep{SchmidtHieber2020Nonparametric,KohlerLanger2021Rate,TangYang2023Minimax,ChaiFanGu2026NDRE}.

\subsection{Asymptotic Theory}

We now present our main theoretical results that show the consistency and asymptotic normality of our DA estimator with the root-$n$ rate.
\begin{proposition}[Consistency of DA]\label{prop:DA-consistency}
Under Assumptions \ref{Assump:ID} to \ref{Assump:first-step-L2}, we have \(\thetahat\to_{\Pp}\theta_0\). \end{proposition}

\begin{theorem}[Asymptotic Normality of DA]\label{thm:DA-AN}
Suppose Assumptions \ref{Assump:ID} to \ref{Assump:Moments} hold. Then, we have the following asymptotic representation
\begin{equation}
\sqrt n(\thetahat-\theta_0)
=-H^{-1}
\left[
   \G_n\{s_0-b\varsigma_0\}
   +b\sqrt r\,\G_N\varsigma_0
\right]+\op(1),
\label{eq:DA-linear-representation}
\end{equation}
where \(b=(1+r)^{-1}\). Consequently, our estimator is asymptotically normal
\[
\sqrt n(\thetahat-\theta_0)\rightsquigarrow N(0,\Sigma_r),
\]
with the asymptotic covariance matrix $\Sigma_r=H^{-1}\Omega_r (H^{-1})^{\top}$, where
\begin{align}
\Omega_r=\Var\{s_0-b\varsigma_0\}+b^2r\Var(\varsigma_0).
\label{eq:Omega-simplified}
\end{align}
\end{theorem}
\begin{remark}[Information in Surrogates]
The asymptotic variance itself reveals an interesting observation about the quality of the surrogate variable $\widehat{Y}$. Considering the asymptotic variance formula, 
\[
 \Cov(s_0,\varsigma_0)
 =\E\{\E(s_0\mid V)\varsigma_0^\top\}
 =\Var(\varsigma_0),
\]
which shows $\Sigma_r=\Var(s_0)-\frac{1}{1+r}\Var(\varsigma_0)$. Suppose the equivalence between two \textit{sigma-algebras} \(\sigma(X,\widehat Y)=\sigma(X,Y)\), so that the information generated by $(X,\widehat{Y})$ can perfectly recover the information in $\sigma(X,Y)$, then we have $s_0\equiv \varsigma_0$. The asymptotic variance of $\sqrt{n}(\thetahat-\theta_0)$ becomes $\frac{r}{1+r}H^{-1}\Var(s_0) (H^{-1})^{\top}$. Thus, if we choose the normlization based on the entire data, then we have $\sqrt{n+N}(\thetahat-\theta_0)\rightsquigarrow N(0,H^{-1}\Var(s_0) (H^{-1})^{\top})$. This achieves the same oracle asymptotic variance, as if we have observed $Y_i$ from the unlabeled data and constructed the \(M\)-estimator using the entire data set.   
\end{remark}

\begin{remark}[Efficiency Issues]
The covariance in \eqref{eq:Omega-simplified} agrees with the most efficient one in the literature of PPI \citep{ji2025PredictionsSurrogates}. We want to clarify the efficiency stemmed from our theorem. Although we start with the \(M\)-estimation problem, there is a correponding \(Z\)-estimation (at least for smooth criterion functions) with its score function or estimating equation. The fully semiparametric efficient estimator in \cite{RobinsRotnitzkyZhao1994MissingRegressors} needs to start from the optimal score function or estimating equation (by choosing the optimal $h^*$ defined therein). This is a well understood problem. For instance, for the linear regression model with errors having zero conditional mean, one should use the generalized least squares with nonparametrically estimated conditional heteroskedasticity \citep{Robinson1987Hetero}. For the quantile regerssion, the default Koenker-Basset (K-B) estimator \citep{KoenkerBassett1978Quantile} also fails to achieve the semiparametric efficiency and one has to use conditional density weighted estimation approach \citep{NeweyPowell1990Censored}. Nonetheless, those highly sophisticated (and semiparametrically efficient) options do not match the popularity of OLS or K-B estimation in applied works. Furthermore, the determination of the score is not even in closed-form for general estimating equations. In the context of semi-supervised learning, which is closely related to the PPI literature, \cite{xu2025UnifiedFrameworkSemiparametricEfficiency} propose the one-step updated version that makes use of the efficient score from the labeled data. Our purpose is to make the best use of the unlabeled data for a given \(M\)-estimation problem. In other words, our efficiency is stated for a given $h$ function in \cite{RobinsRotnitzkyZhao1994MissingRegressors}. If the corresponding loss function attached to the efficient score is indeed available, one can still apply our method and achieve the full semiparametric efficiency.      
\end{remark}

When it comes to the theory of our DDA, we focus on its asymptotic normality and its corresponding weaker regularity conditions, compared with our DA. Regarding the root-$n$ consistency property of DDA, one can resort to \cite{ji2025PredictionsSurrogates} to construct the linearized correction term to keep the convexity of the sample criterion function. Alternatively, we can assume that DDA approximately solves the sample score function $\partial_{\theta}\widehat{Q}_{DDA}(\thetabc)=o_p(n^{-1/2})$. Then, one can utilize the existing theory from the two-step \(Z\)-estimation to derive the root-$n$ convergence rate. See \cite{ChenLintonVanKeilegom2003Nonsmooth} and \cite{NanWellner2013SemiparametricZ}.

\begin{theorem}[Asymptotic Normality of DDA]\label{thm:DDA-AN}
We assume that the debiased distributional aligned estimator is root-$n$ consistent. Suppose Assumptions \ref{Assump:ID} to \ref{Assump:SE}, \ref{Assump:Moments}, and \ref{Assump:DDA-local} hold.  Then, the following asymptotic representation holds
\begin{equation}
\sqrt n(\thetabc-\theta_0)
=-H^{-1}
\left[
   \G_n\{s_0-b\varsigma_0\}
   +b\sqrt r\,\G_N\varsigma_0
\right]+\op(1),
\label{eq:DDA-linear-representation}
\end{equation}
and therefore \(\sqrt n(\thetabc-\theta_0)\rightsquigarrow N(0,\Sigma_r)\), with the same asymptotic variance \(\Sigma_r\) given in Theorem \ref{thm:DA-AN}. 
\end{theorem}

\begin{remark}[Distributional Difference between Labeled and Unlabeled Data]\label{remark:DistDiff}
The common distribution of the labeled and unlabeled data in \cref{Assump:Sampling} is stronger than necessary. For point identification of $\theta_0$ from Equation \eqref{eq:eq12}, it suffices to assume that 
\begin{itemize}
    \item[(i)] 
the support of $(X^U,\widehat{Y}^U)$ is contained in the support of $(X^L,\widehat{Y}^L)$, 
\item[(ii)] $
F_0^U(y\mid X^U,\widehat Y^U)=F_0^L(y\mid X^U,\widehat Y^U)$, and 
\item[(iii)] $\mathbb E[\ell_\theta(X^U,Y^U)]=\mathbb E[\ell_\theta(X^L,Y^L)]$ for all $\theta\in\Theta$. 
\end{itemize}    

In applications where the parameter of interest is associated with the unlabeled population, i.e.,
\[
\theta_0:=\arg\min_{\theta\in\Theta}\E\bigl[\ell_{\theta}(X^U,Y^U)\bigr],
\]
(iii) above can be discarded. In other words, the labeled and unlabeled samples can have different marginal distributions. This is the general auxiliary-sample case considered in Section 2.2.1 of \citet{chen2005MeasurementErrorModels}. Under conditions (i) and (ii) above, 
\begin{equation}
\theta_0
=\arg\min_{\theta\in\Theta}
\E\left[\int \ell_{\theta}(X^U,y)\, dF_0\bigl(y\mid X^U,\widehat{Y}^U\bigr)\right].
\end{equation}
This perspective helps motivate the AEM approach of \citet{zhang2025AgenticEconomicModeling}, which uses labeled data primarily to estimate the conditional distribution.
\end{remark}

\subsection{Comparison of DA and DDA}
\subsubsection{Verification of  Assumption \ref{ass:first-step-AL}}
The crucial assumption that leads to the asymptotic representation of our DA estimator is Assumption \ref{ass:first-step-AL}. As we show in the proof of Theorem \ref{thm:DA-AN}, this stems from analyzing certain linear functional of the estimated conditional distribution. For classical kernel smoothing and sieve type estimators, this can be directly shown\footnote{For the local constant estimator, we refer to \cite{WangChen2009EmpiricalLikelihoodMissing} for the proof of $\check{T}_{n1}$ term therein. For the sieve estimator, we refer to the proof of equation (A.2.0) in \cite{chen2005MeasurementErrorModels}. For completeness, we also provide a direct proof for the local linear estimator in Subsection \ref{subsec:small-bias-kernel-series}, which is used in our Monte Carlo simulation.}.

For other possible first-stage conditional distribution estimator, we demonstrate that the key to Assumption \ref{ass:first-step-AL} is the no bias condition \citep{vdVaart1998AsympStat}, well established in semiparametric estimation problems. Define this functional for any conditional distribution $F$:
\begin{equation}
	\Psi(F):=P_V\!\left[\int s_0(X,y)\,dF(y\mid V)\right].
	\label{eq:Psi-F}
\end{equation}
We further define the residual score as $\phi_F(Z):=s_0(Z)-\varsigma_{\theta_0}(V;F)$. At the truth, we write it as
\begin{equation}
	\phi_0(Z)=s_0(Z)-\varsigma_0(V).
	\label{eq:conditional-IF}
\end{equation} 
Note that $\mathbb{E}[\phi_0(Z)|V]=0$ by construction.

The expansion that characterizes the first-step estimation effect is 
\begin{equation}
   P_V\Delta_n(\cdot;\widehat F)= \Psi(\widehat F)-\Psi(F_0)=-P_0\phi_{\widehat F},
\end{equation}
which holds because \(\Psi(F_0)=0\). Then, we have the following identity
\begin{align}	
	\Psi(\widehat F)-\Psi(F_0)=(\Pn-P_0)\phi_{0}+(\Pn-P_0)(\phi_{\widehat F}-\phi_{0})-\Pn\phi_{\widehat F}
	\label{eq:correct-sign-decomp}
\end{align}
Hence, if we assume that
\begin{align}
	\Pn\phi_{\widehat F}&=\op(n^{-1/2}),\label{eq:score-calibration-main}\\
	\G_n(\phi_{\widehat F}-\phi_0)&=\op(1),\label{eq:IF-equicontinuity-main}
\end{align}
then
\[
\sqrt n\{\Psi(\widehat F)-\Psi(F_0)\}
=\G_n\phi_0+\op(1)=\G_n\{s_0-\varsigma_0\}+o_p(1),
\]
which is exactly the asymptotic representation in Assumption \ref{ass:first-step-AL}. The second condition in (\ref{eq:IF-equicontinuity-main}) boils down to check the stochastic equicontinuity of the functional class $\{\varsigma_{\theta_0}(\cdot;F):F\in\mathcal{F}\}$. The first condition (\ref{eq:score-calibration-main}) is the well-known no-bias condition \citep{vdVaart1998AsympStat}\footnote{Besides an explicit debiasing as in our DDA, another possibility is the targeted learning approach, which further calibrates the conditional distribution estimator defined by $\widetilde{F}$, so that $\Pn\phi_{\widetilde F}=\op(n^{-1/2})$. The book-length treatment can be found in \cite{vdLaanRose2011Targeted}}. 

We can further characterize the generalized linear models. Suppose for the moment that $Y$ is scalar and, in a neighborhood of $\theta_0$, the score is affine in $Y$,
\begin{equation}
	s_\theta(X,Y)=a_\theta(X)-B(X)Y,
	\qquad B(X)\in\mathbb R^p.
	\label{eq:affine-score-in-Y}
\end{equation}
Let the conditional expectation be denoted by
\begin{equation}
	\nu_0(V):=\E[Y\mid V],
	\qquad
	\widehat\nu(V):=\int y\,d\widehat F(y\mid V).
	\label{eq:nu-def}
\end{equation}
Then, we have $\varsigma_{\theta}(V;\widehat F)=a_\theta(X)-B(X)\widehat\nu(V)$ and $\Delta_n(V;\widehat{F})=-B(X)\{\widehat\nu(V)-\nu_0(V)\}.$ Considering \eqref{eq:score-calibration-main} and \eqref{eq:IF-equicontinuity-main}, Assumption \ref{ass:first-step-AL} follows if
	\begin{align}
		\sqrt n\,\Pn\!\left[B(X)\{\widehat\nu(V)-Y\}\right]&=\op(1),
		\label{eq:score-calibration-affine}\\
		\G_n\!\left[B(X)\{\widehat\nu(V)-\nu_0(V)\}\right]&=\op(1).
		\label{eq:score-stability-affine}
	\end{align}
We focus on the no bias condition in \eqref{eq:score-calibration-affine}. In the example of estimating the mean of outcome, it reduces exactly to $\sqrt n\,\Pn\{\widehat\nu(V)-Y\}=\op(1)$.
In the example of linear regression, it is equivalent to $\sqrt n\,\Pn\!\left[X\{\widehat\nu(V)-Y\}\right]=\op(1)$. Hence, the no-bias condition states that the residual $(Y_i-\widehat{\nu}(V_i))_{i=1}^n$ has (approximately) zero sample mean, or it is approximately orthogonal to the regressors $X$. This condition is particularly natural for the linear sieve estimator, if one incorporates $X$ in the basis functions of estimating the conditional mean $\widehat{\nu}(\cdot)$. This is also related to the idempotent property of the sieve method discussed by \cite{NeweyHsiehRobins2004Twicing}.

\subsubsection{Local Robustness and Stochastic Equicontinuity}
For a highly complicated \(\widehat F\), we instead impose Assumption \ref{Assump:DDA-local} for our DDA estimator. Note that the left hand side of Equation (\ref{eq:DDA-local-orthogonality}) can be expressed as the difference of two empirical process terms. If the local linearization condition in Assumption (\ref{Assump:PlugLocal}) holds for both $\QN$ and $\Pn$, then we can bound 
\begin{align}
	&\sup_{\|h\|\le M}n\left|(\QN-\Pn)D_{n,h}(\cdot;\widehat{F})\right| \notag \\
    \leq &\sup_{\|h\|\le M}\sqrt{n}\left|(\QN-P_V)h^\top\Delta_n(\cdot;\widehat{F})-(\Pn-P_0)h^\top\Delta_n(\cdot;\widehat{F})\right|+o_p(1),\label{inequality:bound}
\end{align}
because the same centering term $P_V\Delta_n(\cdot;\widehat{F})=P_0\Delta_n(\cdot;\widehat{F})$ cancels out. The right hand side of the inequality can be shown to be asymptotically negligible under the stochastic equicontinuity on the $\Delta_n(\cdot;\widehat{F})$, which can be deduced from proper Donsker property \citep{vdVaartWellner2023EP}. This reveals the robustness of DDA or the related PPI type construction with respect to the asymptotic properties of the first step estimation of the conditional distribution. In addition, it also introduces the debiasing step to the kernel imputation method in the missing data \citep{WangChen2009EmpiricalLikelihoodMissing}, as well as the measurement error model with validation data (assuming the same marginal distribution) \citep{chen2005MeasurementErrorModels}. In Section \ref{Sec: DDA and DeepLearning}, we showcase the verification of the stochastic equicontinuity for deep neural network (DNN) based estimators, building on the recent development of \cite{SchmidtHieber2020Nonparametric,KohlerLanger2021Rate,ChaiFanGu2026NDRE}.

\subsection{A Multiplier Bootstrap and Its Consistency}\label{subsec:bootstrap}

We construct a multiplier bootstrap for inference on $\theta_0$, directly motivated from the asymptotic linear representation
of our estimators. In our construction, the fitted conditional distribution is held fixed throughout the bootstrap,
and its estimation effect is reproduced by multiplier bootstrap weights together with the estimated influence
function.  In comparison, a naive bootstrap that re-optimizes the sample criterion would generally require re-estimating the first-step conditional distribution (and potentially re-running conditional simulation) in every bootstrap replication. 

Bootstrap inference for semiparametric estimators has been studied extensively in the literature.
\citet{WellnerZhan1996BootstrapZ} establish a general bootstrap theorem for
possibly infinite-dimensional $Z$-estimators.
\citet{ChenLintonVanKeilegom2003Nonsmooth} and \citet{MaKosorok2005WeightedBootstrap} develop nonparametric and weighted bootstrap theory for
semiparametric $M$-estimation, respectively. \citet{ChengHuang2010Bootstrap} establish
bootstrap consistency under exchangeable weights. More recently,
\citet{TangWestling2024Bootstrap} study asymptotically linear estimators given nuisance functions estimated by complex machine learning methods, and \citet{LinHan2026BootstrapDML}
justify exchangeably weighted bootstraps for cross-fitted DML estimators.

Let \(\widehat\theta\) denote either estimator in our proposal. Define
\[
    \widehat s_i:=s_{\widehat\theta}(X_i,Y_i),
    \qquad
    \widehat \varsigma_i:=\varsigma_{\widehat\theta}(V_i;\widehat F),
    \qquad
    \widehat v_i:=\widehat s_i-b_n\widehat \varsigma_i.
\]
Referring to bootstrap weights, $\xi_1,\ldots,\xi_n$ are \textit{i.i.d.} from some distribution $F_{\xi}$, and $\zeta_{n+1},\ldots,\zeta_{n+N}$ are \textit{i.i.d.} from
a (possibly different) distribution $F_{\zeta}$. 
Consider a consistent estimator of the Hessian matrix \(\widehat H\to_{\Pp}H\) \citep{Chen2007Sieve}. In practice, we can make use of the numerical derivative based on the sample loss function computed over the labeled data. Set the bootstrap statistic by
\begin{align}
T_n^*
:=
-\widehat H^{-1}
\Bigg[
\frac1{\sqrt n}\sum_{i=1}^n
  (\xi_i - \bar{\xi})\widehat v_i
+
b_n\sqrt{r_n}\,
\frac1{\sqrt N}\sum_{j=n+1}^{n+N}
  (\zeta_j - \bar{\zeta}) \widehat{\varsigma}_j
\Bigg].
\end{align}
where $\bar{\xi} = n^{-1}\sum_{i=1}^n \xi_i$ and $\bar{\zeta} = N^{-1}\sum_{i=n+1}^{n+N} \zeta_i$.

We introduce a collection of high-level assumptions that lead to the bootstrap consistency theory, followed by some heuristic discussions.
\begin{assumption}[Bootstrap Conditions]
\label{ass:bootstrap-feasibility}
The following conditions hold:
\begin{enumerate}
\renewcommand{\labelenumi}{(\roman{enumi})}
\item The estimator for the Hessian matrix satisfies
$\|\widehat H-H\|_{\mathrm{op}}=o_p(1)$.
\item The fitted score functions satisfy:
\begin{align}
\frac1n\sum_{i=1}^n
 \|\widehat s_i-s_0(Z_i)\|^2&=o_p(1),
~~\frac1n\sum_{i=1}^n
 \|\widehat\varsigma_i-\varsigma_0(V_i)\|^2=o_p(1),
\label{eq:bootstrap-empirical-varsigma-L}\\
\frac{r_n}{N}\sum_{j=n+1}^{n+N}
 \|\widehat\varsigma_j-\varsigma_0(V_j)\|^2&=o_p(1).
\label{eq:bootstrap-empirical-varsigma-U}
\end{align}
\item Two vectors $(\xi_1,\ldots,\xi_n)$ and $(\zeta_{n+1},\ldots,\zeta_{n+N})$
are mutually independent and jointly independent of the observations $\{Z_i\}_{i=1}^n$ and
$\{V_j\}_{j=n+1}^{n+N}$. For every $i=1,\ldots,n$ and
$j=n+1,\ldots,n+N$,
\begin{equation}
 \E\xi_i=\E\zeta_j=0,\qquad
 \E\xi_i^2=\E\zeta_j^2=1.
\label{eq:bootstrap-multiplier-moments}
\end{equation}
\end{enumerate}
\end{assumption}
Assumption (i) is standard for the Hessian matrix in two-step semiparametric estimation. Conditional on the observed data,
the statistic $T_n^*$ can be written as a sum with independent bootstrap weights. The consistency in the empirical $L_2$-norm stated in Assumption \ref{ass:bootstrap-feasibility} (ii) controls the estimation error of $\widehat{\theta}$ and $\widehat{F}$. This can be further checked by proper continuity on $\Vert s_{\theta}-s_0\Vert^2_{P_0,2}$ and $\Vert \varsigma_{\theta}(\cdot;F)-\varsigma_{\theta_0}(\cdot;F)\Vert^2_{P_0,2}$, as well as imposing Glivenko-Cantelli properties on those classes. As we work with independent bootstrap weights, it suffices to assume finite second moments as in Assumption \ref{ass:bootstrap-feasibility} (iii), in line with \cite{MaKosorok2005WeightedBootstrap}. For general exchangeable weights, stronger moments are required \citep{WellnerZhan1996BootstrapZ,ChengHuang2010Bootstrap}. 
Let $\mathrm{BL}_1(\mathbb R^p)$ denote the real functions bounded in
absolute value by one with Lipschitz constant at most one \citep{vdVaartWellner2023EP}.
\begin{theorem}[Bootstrap Consistency]\label{Thm:Bootstrap}
Suppose that the conditions of Theorem~\ref{thm:DA-AN} hold for
$\widehat\theta=\thetahat$, or the conditions of Theorem~\ref{thm:DDA-AN} hold for
$\widehat\theta=\widehat\theta_{DDA}$. Maintain
Assumption~\ref{ass:bootstrap-feasibility}. Then, we have
\begin{align}
\sup_{f\in\mathrm{BL}_1(\mathbb R^p)}
 \big|\E^*f(T_n^*)-
 \E f\{\sqrt n(\widehat\theta-\theta_0)\}\big|\to_p0.
\label{eq:bootstrap-sampling-approx}
\end{align}

\end{theorem}

The above theorem implies that conditionally on the observed data,
$T_n^*\rightsquigarrow_{\Pp}N(0,\Sigma_r)$, where \(\Sigma_r\) is the asymptotic covariance matrix of\(\sqrt n(\widehat\theta-\theta_0)\) by Theorem 1.13.1 of \cite{vdVaartWellner2023EP}. Let $\theta_{0,j}$ be the $j$th coordinate of $\theta_0$, and let $\widehat{\theta}_j$ be the corresponding estimator. A $(1-\alpha)\times 100\%$ confidence interval for $\theta_{0,j}$ can be formed as
\begin{align*}
    \mathcal{C}_{n,j}(\alpha)
    =\left[\widehat{\theta}_{j} - Q^*_{n,j}(1-\alpha/2)/\sqrt{n},\ \widehat{\theta}_{j} - Q^*_{n,j}(\alpha/2) /\sqrt{n} \right],
\end{align*}
where $Q^*_{n,j}(a)$ denotes the $a$-th quantile of the bootstrap distribution $\{\widetilde{T}^*_{n,b,j}: b=1,\ldots,B\}$, and $\widetilde{T}^*_{n,b,j}$ is the $j$th coordinate of the $b$th draw from $T_n^*$. An important consequence of Theorem \ref{Thm:Bootstrap} is that, for each fixed $j\in\{1,\ldots,p\}$,
\begin{equation}
P_0\bigl\{\theta_{0j}\in \mathcal{C}_{n,j}(\alpha)\bigr\}\to 1-\alpha.
\end{equation}

\section{Comparisons with Existing Proposals}\label{sec:comparisons}
Two recent lines of work address the use of surrogate outcomes: prediction-powered inference (PPI) and agentic economic modeling (AEM). In this section, we compare our methodology with each approach. In addition, we discuss the connection of our approach to the classical results \citep{chen2005MeasurementErrorModels} for measurement error models with validation data.

\subsection{Prediction-Powered Inference and Recalibrated PPI}\label{subsec:PPI}
A generic PPI estimator for $M$-estimation can be written in our notation as
\begin{equation}
  \widehat\theta_g^{\mathrm{PPI}}
  \in\arg\min_{\theta\in\Theta}
  \left\{
     \Pn\ell_\theta+(\QN-\Pn)g_\theta
  \right\},
  \label{eq:PPI-g-current}
\end{equation}
where $g_\theta=g_\theta(V)$ is an augmented loss. When $g_\theta$ is fixed or estimated using suitable sample-splitting techniques, the augmentation term $(\QN-\Pn)g_\theta$ has population mean zero. The original PPI \citep{angelopoulos2023PredictionpoweredInference} takes $g_\theta=\ell_\theta$, which yields
\begin{equation}
 \widehat\theta_{\ell}^{\mathrm{PPI}}
 \in\arg\min_{\theta\in\Theta}
 \left\{
 \Pn\ell_\theta
 +(\QN-\Pn)\ell_\theta
 \right\}.
 \label{eq:standard-PPI-current}
\end{equation}
Although this correction removes the population bias that would generally arise from substituting the surrogate outcome for the true outcome in the unlabeled sample, the original PPI does not necessarily improve efficiency relative to labeled-only $M$-estimation $\widehat{\theta}_M$. This motivates subsequent work on efficient choices of the augmentation function $g_\theta$; see \citet{angelopoulos2024PPIEfficientPPI}.

\citet{ji2025PredictionsSurrogates} show that an optimal augmented loss satisfies
\begin{equation}
    \partial_\theta g_{\theta_0}^*(V)
    =\frac{1}{1+r}\E[s_0(Z)\mid V].
    \label{eq:optimal-PPI-gradient-current}
\end{equation}
Their RePPI implementation first forms an initial estimator $\widehat\theta_I$, learns a conditional score function $ \widehat s_I(V)
    = \widehat{\E}\bigl[s_{\widehat\theta_I}(Z)\mid V\bigr]$, and then uses a linearized imputed loss,
\begin{equation}
    \widehat g_\theta^{\mathrm{RePPI}}(V)
    =b_n\,\theta^\top\widehat M\widehat s_I(V).
    \label{eq:RePPI-linear-loss-current}
\end{equation}
which preserves convexity. The matrix $\widehat M$ is an adjustment term that affects efficiency; the optimal choice requires inverting the covariance matrix of the score function (cf. Equation (13) in \citealt{ji2025PredictionsSurrogates}).

In contrast, our DA estimator estimates the full conditional distribution $F_0(\cdot\mid V)$ and uses the integrated loss $\mu_{\theta}(\cdot;\widehat F)$, without explicitly subtracting any augmented loss term. Theorem \ref{thm:DA-AN} yields the same influence function and attains the same efficiency bound as the optimal PPI methodology. Structural econometrics examples \citep{HortacsuJoo2023SE} highlight another advantage of our approach. In many applications, the score function itself can be difficult to compute.

Admittedly, DA requires the first-stage conditional distribution estimator to admit a tractable asymptotic linear representation, i.e., the relevant linear functional must satisfy \eqref{eq:first-step-AL}. To facilitate applications with highly adaptive conditional distribution estimators (including deep learning), we propose the locally robustified version,
\begin{equation}
    \widehat Q_{\mathrm{DDA}}(\theta)
    =\Pn\ell_\theta+(\QN-\Pn)g_\theta^{\mathrm{DDA}},
    \qquad
    g_\theta^{\mathrm{DDA}}(V)
    :=b_n\mu_\theta(V;\widehat F).
    \label{eq:DDA-is-PPI}
\end{equation}
Thus, DDA can be viewed as a PPI-type estimator. Subtracting the labeled integrated loss removes the first-order effect of estimating the conditional distribution and yields the local robustness condition in Assumption \ref{Assump:DDA-local}. This is in line with doubly robust estimation \citep{RobinsRotnitzkyZhao1994MissingRegressors} and the DML literature \citep{ChernozhukovEtAl2018DML,ChernozhukovEtAl2022LocallyRobust}.

\subsection{Agentic Economic Modeling in Conjoint Analysis}
Our new perspective provides a cohesive framework linking the PPI literature with semiparametric two-step estimation, which serves as a benchmark for comparing alternative methodologies such as the Agentic Economic Modeling (AEM) approach \citep{zhang2025AgenticEconomicModeling}\footnote{Our discussion mainly concerns its core in econometrics. AEM also contains additional agentic components, including persona generation and mixture constructions, that are outside our framework.}. The recent AEM of \citet{zhang2025AgenticEconomicModeling} follows a Generation--Correction--Inference pipeline. In its conjoint application, raw LLM choices and task features are used with a smaller primary human sample to learn a so-called correction model for the conditional human choice distribution.  In our notation, this correction step is precisely the estimation of   $F_0(\,\cdot\mid V)$, with $V=(X,\widehat Y)$, where $\widehat Y$ denotes the raw LLM-generated choice. 

The connection is especially informative for multinomial choice models in conjoint analysis from Section 4 of \cite{zhang2025AgenticEconomicModeling}. Let $Y\in\{1,\ldots,J\}$ and let $p_{\theta,j}(X)$ be the model-implied probability of choice $j$. With negative log-likelihood loss
$\ell_\theta(X,Y)=-\sum_{j=1}^J\mathbf 1\{Y=j\}\log p_{\theta,j}(X)$,
write $\widehat p_j(V):=\widehat F(Y=j\mid V)$. The integrated loss is then
\begin{equation}
    \mu_\theta(V;\widehat F)
    =-\sum_{j=1}^J\widehat p_j(V)\log p_{\theta,j}(X),
    \label{eq:AEM-soft-loss}
\end{equation}
which is in line with the K-L divergence in Section 4.3 of \cite{zhang2025AgenticEconomicModeling}. The original version of \citet{zhang2025AgenticEconomicModeling} uses the unlabeled data in the second-stage optimization, which can be represented\footnote{See Appendix A.2 of \cite{zhang2025AgenticEconomicModeling} about their definition of $\beta^{pre-trained}$. Given the fitted conditional distributions/choice probabilities, they also use the auxiliary data (our unlabeled data) for estimating the parameter of interest.} abstractly as
\begin{equation}
    \widehat\theta_{\mathrm{AEM}}
    \in\arg\min_{\theta\in\Theta}\;\QN\mu_\theta(\,\cdot\,;\widehat F).
    \label{eq:AEM-abstract-current}
\end{equation}

Our DA criterion differs from \eqref{eq:AEM-abstract-current} and clarifies the asymptotic role of labeled data. DA retains the labeled data in constructing the loss function:
\begin{equation}
    \widehat Q_{\mathrm{DA}}(\theta)
    =a_n\Pn\ell_\theta+b_n\QN\mu_\theta(\,\cdot\,;\widehat F).
    \label{eq:DA-vs-AEM-current}
\end{equation}
 Theorem \ref{thm:DA-AN} implies that $\widehat\theta_{\mathrm{AEM}}$ is asymptotically efficient only when $r=0$. This corresponds to the primary applied scenario emphasized by \citet{zhang2025AgenticEconomicModeling}, where the labeled sample is negligible relative to the unlabeled one. To the best of our knowledge, this econometric perspective on AEM is new.

\subsection{Optimally Weighted GMM}

Our modeling framework is closely related to the classical validation sample design in measurement error problems. See \citet{chen2005MeasurementErrorModels, chen2008SemiparametricEfficiencyGMM}. In that literature, one observes an error-prone proxy for a latent target (either an outcome or covariate) in a primary sample which corresponds to our unlabeled sample, together with a smaller validation sample (our labeled sample) where the latent target is observed. In addition to conceptual overlap, we provide a more careful analysis that connects our DA estimator to the efficient GMM estimator in \cite{chen2005MeasurementErrorModels}. We emphasize that the framework of \cite{chen2005MeasurementErrorModels} is general so that it allows for the over-identified moment conditions (even with the labeled data only) and different marginal distributions of covariates, whereas we focus on the \(M\)-estimation with the identical distribution herein\footnote{\cite{DuanPelger2026AI} consider the GMM setting in which the moment condition from the labeled data alone is over-identified, and they propose PPI-type corrections. They also study certain data-adaptive labeling designs when human labels are added where the 
AI-generated features are less reliable.}.

Recall that our formulation in Equation \eqref{eq:eq12}. For simplicity, we maintain the smoothness on the loss function with respect to the parameter $\theta$. This leads to the following moment condition when we simply add the labeled and unlabeled parts together:
\begin{equation}
\mathbb E\,s_{\theta_0}(X,Y)+\mathbb E\underbrace{\int s_{\theta_0}(X,y)\, dF_0(y\mid X,\widehat Y)}_{ \varsigma_{\theta}(V;F_0)}=0.
\label{eq:DAMM}
\end{equation}
Notice that we can stack up the two sets of moment conditions as in \cite{chen2005MeasurementErrorModels}. Under Assumptions \ref{Assump:Sampling} and \ref{Assump:ID}, $\theta_0$ satisfies the following moment condition:
\begin{align*}
    \mathbb{E}\Biggl[\begin{pmatrix}
        s_{\theta_0}(Z) \\
        \varsigma_{\theta_0}(V;F_0)
    \end{pmatrix}\Biggr] = 0.
\end{align*}
A two-step GMM estimator of $\theta_0$ with the optimal weighting matrix is given by 
\begin{align*}
    \widehat{Q}_{\mathrm{GMM}}(\theta)
    =
    \begin{pmatrix}
     \mathbb{P}_n s_{\theta} \\ 
     \mathbb{Q}_N \varsigma_{\theta}(\cdot;\widehat{F})
    \end{pmatrix}^{\!\top}
    \widehat{W}^{-1}
    \begin{pmatrix}
     \mathbb{P}_n s_{\theta} \\ 
     \mathbb{Q}_N \varsigma_{\theta}(\cdot;\widehat{F})
    \end{pmatrix},
\end{align*}
where $\widehat{W}$ is a consistent estimator of the optimal weighting matrix $W_{\mathrm{GMM}}$:
    \begin{align*}
        W_{\mathrm{GMM}} & = \begin{pmatrix}
            I & 0 \\
            I & I
        \end{pmatrix}
        \begin{pmatrix}
            \mathrm{Var}(s_0) & - \mathrm{Cov}(s_0, \varsigma_0) \\
            -\mathrm{Cov}(s_0, \varsigma_0)^{\top} & (1+r) \mathrm{Var}(\varsigma_0)
        \end{pmatrix}
        \begin{pmatrix}
            I & I \\
            0 & I
        \end{pmatrix} \\
        & =
        \begin{pmatrix}
            I & 0 \\
            I & I
        \end{pmatrix}
        \begin{pmatrix}
            \mathrm{Var}(s_0) & - \mathrm{Var}(\varsigma_0) \\
            -\mathrm{Var}( \varsigma_0) & (1+r) \mathrm{Var}(\varsigma_0)
        \end{pmatrix}
        \begin{pmatrix}
            I & I \\
            0 & I
        \end{pmatrix}
        \end{align*}
Let $\widehat{\theta}_{\mathrm{GMM}}$ denote the two-step estimator. 
Following \citet[Section 3.4]{chen2005MeasurementErrorModels}, one can show that under regularity conditions stated therein, $\widehat{\theta}_{\mathrm{GMM}}$ is root-$n$ consistent and it has the same asymptotic distribution as our DA estimator. 

We reveal a deeper connection of our method to the optimal GMM in \cite{chen2005MeasurementErrorModels}, and vise versa. For the smooth case, we show that our DA estimator approximately solves the first order condition (F.O.C.) of the optimal GMM in \cite{chen2005MeasurementErrorModels}. We decompose the sample criterion of $\widehat{Q}_{GMM}$ into two parts, where the first one is a quadratic term related to our DA estimation, and the second term can be shown that its first order derivative with respect to $\theta$ becomes negligible at $\theta=\widehat{\theta}_{DA}$ or $\theta=\widehat{\theta}_{GMM}$. 

To this end, we introduce some notation. 
Let $\|x\|_A^2 := x^{\top} A x$ for some symmetric matrix $A$ and vector $x$. 
Further,  define
    $T:= \mathrm{Var}(\varsigma_0)$, $M := \mathrm{Cov}(s_0,\varsigma_0)\,\mathrm{Var}(\varsigma_0)^{-1}$, and
    \begin{align*}
        R &:= \mathrm{Var}(s_0) - \frac{1}{1+r}\,\mathrm{Cov}(s_0,\varsigma_0)\,\mathrm{Var}(\varsigma_0)^{-1}\,\mathrm{Cov}(s_0,\varsigma_0)^{\top}.
    \end{align*}
Let $\widehat T$, $\widehat{M}$ and $\widehat{R}$ be consistent estimators of $T$, $M$ and $R$, where $r$ is estimated by $n/N$. 
   
    \begin{proposition} \label{prop:gmm-decomposition}
    Let 
    \begin{align*}
        \widehat{Q}_{\mathrm{GMM}, 1}(\theta)
    &=
    \left\|
    \left(I - \frac{N}{n+N} \widehat{M}\right)
    \mathbb{P}_n s_{\theta}
    +\frac{N}{n+N} \widehat{M} \mathbb{Q}_N \varsigma_{\theta}(\cdot;\widehat{F}) \right\|_{\widehat{R}^{-1}}^2 \mbox{ and}  \\
    \widehat{Q}_{\mathrm{GMM}, 2}(\theta) & = 
    \frac{N}{n+N}
    \left\|
    \mathbb{Q}_N \varsigma_{\theta}(\cdot ;\widehat{F})
    - \mathbb{P}_n s_{\theta}
\right\|_{\widehat{T}^{-1}}^2.
    \end{align*}
    Then, the GMM objective function can be decomposed to two parts as follows: 
    \begin{align*}
        \widehat{Q}_{\mathrm{GMM}}(\theta) = \widehat{Q}_{GMM, 1}(\theta) + \widehat{Q}_{GMM, 2}(\theta).
    \end{align*}
    \end{proposition}

Since $\mathrm{Cov}(s_0, \varsigma_0) = \mathrm{Var}(\varsigma_0)$, we choose $\widehat{M} = I$. Then, $\widehat{Q}_{GMM,1}(\theta) = \|\partial_{\theta} \widehat{Q}_{DA} (\theta)\|^2_{\widehat{R}^{-1}}$,
whose (approximate) F.O.C. is the single combined moment equation
\[
\frac{1}{n+N}\left\{\sum_{i=1}^n s_{\theta}(Z_i)+\sum_{j=n+1}^{n+N}\varsigma_{\theta}(V_j;\widehat F)\right\}=o_p(n^{-1/2}).
\]
This is satisfied for our DA estimator, i.e., $\partial_{\theta} \widehat{Q}_{DA} (\widehat{\theta}_{DA})=o_p(n^{-1/2})$.
At first glance, $\widehat{Q}_{GMM}$ has an additional term $\widehat{Q}_{GMM,2}$. Under some mild regularity conditions, we prove in Proposition \ref{prop:gmm-second-term-negligible} that $\partial_{\theta}\widehat{Q}_{GMM, 2}(\widehat{\theta}_{DA}) = o_p(n^{-1/2})$, which justifies our claim that the DA estimator also approximately solves the F.O.C. of the optimally weighted GMM in \cite{chen2005MeasurementErrorModels}. The discussion following the same Proposition \ref{prop:gmm-second-term-negligible} implies that $\partial_{\theta}\widehat{Q}_{GMM, 2}(\widehat{\theta}_{GMM}) = o_p(n^{-1/2})$ under the set of regularity conditions from \cite{chen2005MeasurementErrorModels}, which also justifies that the optimal GMM estimator in \cite{chen2005MeasurementErrorModels} solves the F.O.C of our DA problem.

\section{Monte Carlo Simulation}\label{sec:numerical}

For illustrative purposes, we consider the following simple regression setup:
\begin{align}
    Y & = \log(1+X)\,\beta_0 + \sigma(X)\,\epsilon, \\
    \widehat{Y} & = X\,\beta_0 + \rho\,\sigma(X)\,\epsilon + \nu,
\end{align}
where $X \sim U(0,1)$, and $\epsilon$ and $\nu$ follow $N(0,1)$ truncated to $(-3,3)$; $X$, $\epsilon$, and $\nu$ are mutually independent. We set $\beta_0 = 1$ and $\sigma(X) = \sqrt{1+(X\beta_0)^2}$. The target model regresses $Y$ on $\log(1+X)$, with squared loss
\[
\ell_{\beta}(X,Y) = \tfrac{1}{2}\bigl(Y - \log(1+X)\,\beta\bigr)^2.
\]
The conditional correlation between $Y$ and $\widehat{Y}$ given $X=x$ equals
$\rho\,\sigma(x)/\sqrt{\rho^2\sigma^2(x)+1}$, so their dependence increases with $|\rho|$. We vary $\rho$ to study the performance of proposed estimators under different strengths of surrogate informativeness.

The numbers of labeled and unlabeled observations are set to $n=3,000$ and $N=7,000$, respectively. We compare AEM, DA, DDA, and DDA with cross-fitting. We estimate the conditional distribution of $Y$ given $(X,\widehat{Y})$ using local linear regression, sieve regression, and MGAN \cite{baptista2024ConditionalSamplingMonotoneGANs}.

We vary $\rho$ from $0$ to large values so that the maximal conditional correlation
\[
\rho_x := \max_{x\in[0,1]} \frac{\rho\sigma(x)}{\sqrt{\rho^2\sigma^2(x)+1}} = \frac{\sqrt{2}\rho}{\sqrt{2\rho^2+1}}
\]
takes values $0, 0.11, 0.22, \ldots, 0.99$. We run 1{,}000 Monte Carlo replications and use 1{,}000 bootstrap draws. For sieve regression we use a B-spline basis with 3 knots and degree 2, where the knots are placed based on quantiles. For local linear regression, we used the tricube product kernel and the bandwidth is set to $h_j \propto n^{-1/3}$ for $j=1, \dotsc, d_V$ where the constant term is chosen by Silverman's rule-of-thumb for simplicity. For MGAN we use an LSGAN objective; the monotonicity penalty is 0.01. The generator and discriminator are multi-layer perceptrons with three hidden layers of 64 units, using leaky ReLU activations (negative slope 0.2). We train with Adam ($\beta=(0.5,0.999)$), initial learning rate $10^{-4}$ with step decay after each batch using factor $\gamma=0.995$, batch size 128, and 300 epochs.

We report bias, standard deviation, and the length and coverage of 95\% bootstrap confidence intervals. Overall, the simulation evidence supports our theory.

\begin{figure}[h]
    \centering
    \includegraphics[width=1\linewidth]{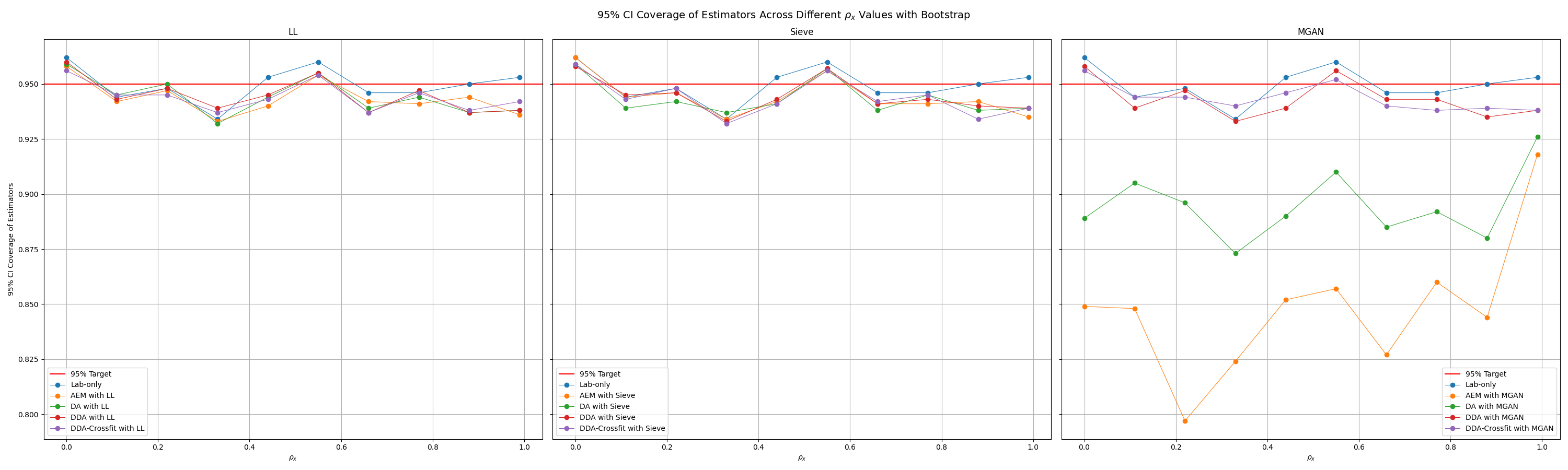}
    \caption{Coverage rates of 95\% bootstrap confidence intervals of several estimators in Monte Carlo simulations, under different estimators of $F(\cdot\mid X,\widehat{Y})$.}
    \label{fig:coverage_boot}
\end{figure}

\begin{figure}[h]
    \centering
    \includegraphics[width=1\linewidth]{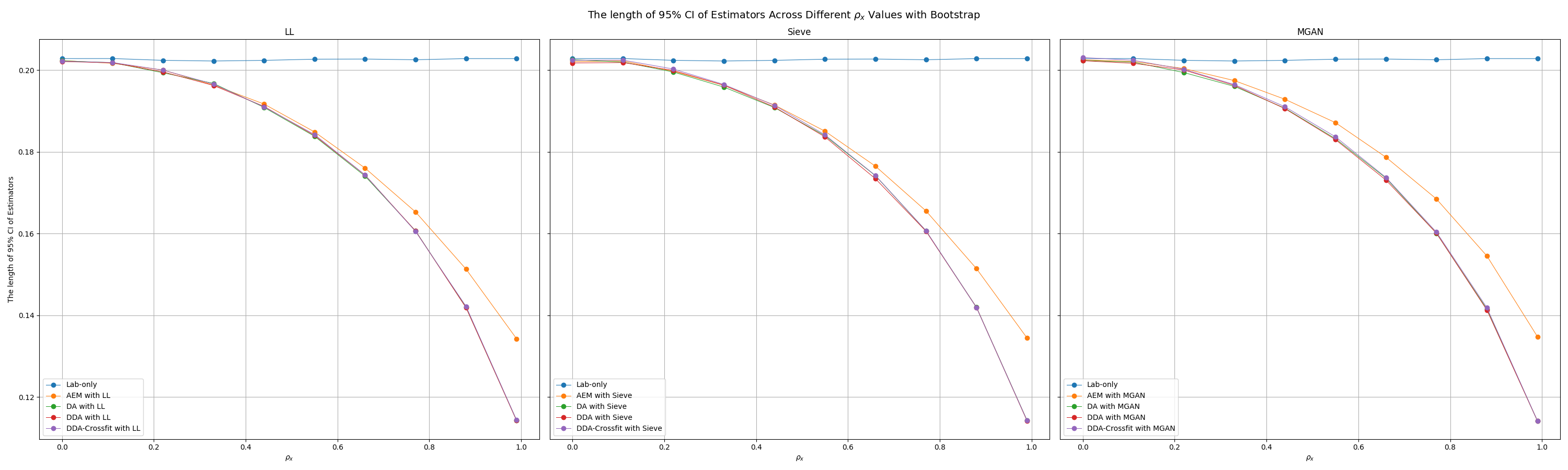}
    \caption{The length of 95\% bootstrap confidence intervals of several estimators in Monte Carlo simulations, under different estimators of $F(\cdot\mid X,\widehat{Y})$.}
    \label{fig:ci_length_boot}
\end{figure}

Figures \ref{fig:coverage_boot} and \ref{fig:ci_length_boot} report the coverage rates and average lengths of 95\% bootstrap confidence intervals. Coverage is close to the nominal 95\% level for most configurations, with the main exceptions occurring when MGAN is used to estimate the conditional distribution. The lengths of 95\% bootstrap confidence intervals for AEM and DA-variants decreases as $\rho$ increases.

\begin{figure}[h]
    \centering
    \includegraphics[width=1\linewidth]{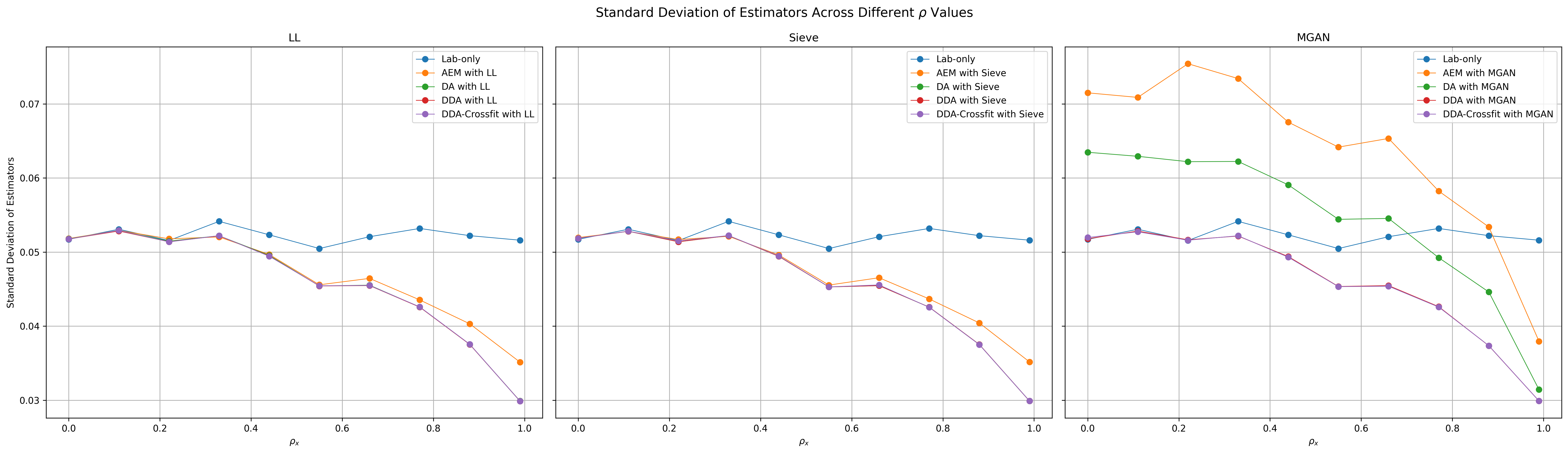}
    \caption{Standard deviations of several estimators in Monte Carlo simulations, under different estimators of $F(\cdot\mid X,\widehat{Y})$.}
    \label{fig:std-comparison}
\end{figure}

\Cref{fig:std-comparison} shows Monte Carlo standard deviations. As $\rho$ (equivalently $\rho_x$) increases, $\widehat{Y}$ becomes more informative and the variability of the AEM and DA-type estimators falls. When $\rho$ is small, these estimators behave similarly to the labeled-only estimator. With sieve estimation of $F(\cdot\mid X,\widehat{Y})$, the AEM and DA standard deviations are especially close to the labeled-only benchmark for small $\rho$, while differences become more pronounced for larger $\rho$. Finally, when MGAN is used in the first stage, applying the locally robust (DDA) correction is important for stable inference, consistent with the coverage patterns.

\begin{figure}[h]
    \centering
    \includegraphics[width=\linewidth]{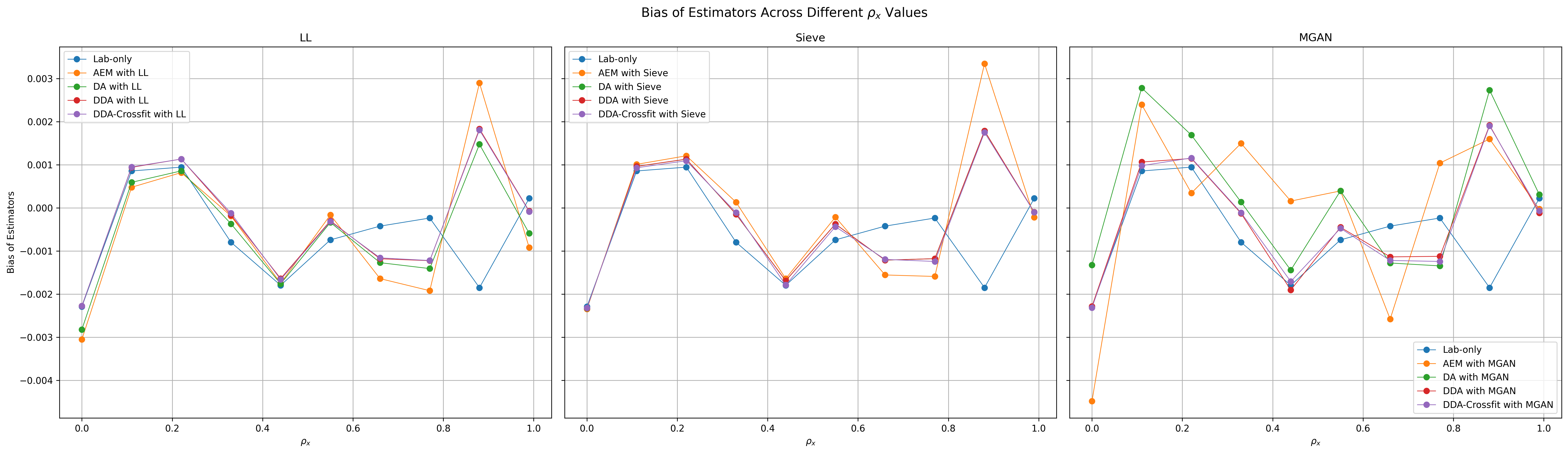}
    
    \caption{Biases of several estimators in Monte Carlo simulations, under different estimators of $F(\cdot\mid X,\widehat{Y})$.}
    \label{fig:bias-comparison}
\end{figure}

\Cref{fig:bias-comparison} shows Monte Carlo biases. In this design the biases are small overall. Sieve regression yields similar (and small) bias across estimators, while local linear regression and MGAN exhibit slightly larger bias for AEM and DA. This is in line with the advocate of sieve methods \citep{NeweyHsiehRobins2004Twicing,chen2005MeasurementErrorModels,Chen2007Sieve} for their small bias properties.

\section{Conclusion}\label{sec:conclusion}
This paper studies the use of externally generated surrogates as auxiliary information when the outcome of interest is observed for only a subset of the sample. We propose a distribution-aligned \(M\)-estimator that incorporates the unlabeled observations through the estimated conditional distribution. We derive its asymptotic distribution accounting explicitly for the first-stage estimation, and show that its asymptotic variance coincides with the efficient PPI benchmark for the given estimating equation. We also propose a local robustification that does not require an asymptotic linear representation for the relevant functional of the estimated conditional distribution. The connections with PPI, AEM and optimally weighted GMM clarify how these procedures can be understood from a unified perspective. For inference, we establish the validity of a multiplier bootstrap that avoids repeatedly estimating the conditional distribution. Monte Carlo simulations illustrate the finite-sample performance of the proposed estimators and support our theory.

\bibliographystyle{apalike}
\bibliography{references_combined}

\appendix

\section{Proof of Main Results}\label{app:proofs}

\begin{proof}[Proof of Proposition \ref{prop:DA-consistency}]
We aim to show the uniform convergence of the sample loss function to the population version for our DA estimator:
	\begin{equation}
		\sup_{\theta\in\Theta}|\widehat Q_{\mathrm{DA}}(\theta)-Q(\theta)|=\op(1).
		\label{eq:ULLN-DA}
	\end{equation}
We proceed with the following decomposition:
\begin{align*}
&\sup_{\theta\in\Theta}|\widehat Q_{\mathrm{DA}}(\theta)-Q(\theta)|\\
\leq & a_n\sup_{\theta\in\Theta}\left|(\mathbb{P}_n-P_0)\ell_{\theta} \right|+b_n\sup_{\theta\in\Theta}\left|(\mathbb{Q}_N-P_V)\mu_{\theta}(\cdot;F_0) \right|\\
+& b_n\sup_{\theta\in\Theta}\left|P_V\left(\mu_{\theta}(\cdot;\widehat F)-\mu_{\theta}(\cdot;F_0) \right) \right|+b_n\sup_{\theta\in\Theta}\left|(\mathbb{Q}_N-P_V)\left(\mu_{\theta}(\cdot;\widehat F)-\mu_{\theta}(\cdot;F_0) \right) \right|.
\end{align*}
The first term on the right hand side of the above inequality is $o_p(1)$ by the Glivenko-Cantelli (G-C) property in Assumption \ref{Assump:GC}. At the true $F_0$, we have the permanence of the G-C property from the original loss $\ell_{\theta}$ extended to the integrated version by Theorem 2.10.5 in \cite{vdVaartWellner2023EP}. Thus, the second term is also $o_p(1)$. The third term is $o_p(1)$ following Assumption \ref{Assump:first-step-L2}. 

For the fourth term, independence of $\widehat F$ (computed over the labeled data) and the unlabeled
sample observations gives
\begin{align*}
    &\E\left[
        \mathbb Q_N\left\{
            \sup_{\theta\in\Theta}
            \left|
                \mu_\theta(\cdot;\widehat F)
                -\mu_\theta(\cdot;F_0)
            \right|
        \right\}
        \,\middle|\,(Z_i)_{i=1}^n
    \right] \\
    =&
    P_V\left[
        \sup_{\theta\in\Theta}
        \left|
            \mu_\theta(\cdot;\widehat F)
            -\mu_\theta(\cdot;F_0)
        \right|
    \right]
    =\op(1).
\end{align*}
An application of (the conditional version) Markov's inequality therefore implies that
\[
    \mathbb Q_N\left[
        \sup_{\theta\in\Theta}
        \left|
            \mu_\theta(\cdot;\widehat F)
            -\mu_\theta(\cdot;F_0)
        \right|
    \right]=\op(1).
\]
Consequently,
\begin{align*}
    &\sup_{\theta\in\Theta}
    \left|
        (\mathbb Q_N-P_V)
        \left\{
            \mu_\theta(\cdot;\widehat F)
            -\mu_\theta(\cdot;F_0)
        \right\}
    \right|\\
    \qquad\leq&
    (\mathbb Q_N+P_V)\left[
        \sup_{\theta\in\Theta}
        \left|
            \mu_\theta(\cdot;\widehat F)
            -\mu_\theta(\cdot;F_0)
        \right|
    \right]
    =\op(1).
\end{align*}

Given the uniform convergence established in (\ref{eq:ULLN-DA}) and the identification condition in Assumption \ref{Assump:ID}, the rest of the proof follows from Theorem 5.7 of \cite{vdVaart1998AsympStat}. 
\end{proof}

Our proof of the asymptotic normality follows the general route by \cite{PakesPollard1989Simulation} and \cite{ChenLintonVanKeilegom2003Nonsmooth}. We first establish a local quadratic expansion of the criterion function and then apply the completion-of-square trick. Notable features in our context include the split between labeled and unlabeled data, as well as the fact that the first-step nonparametric estimation enters linearly into the unlabeled part. We start with the local quadratic expansion. The basic device is to expand the \emph{increment} of a convex loss function. For this analysis, write \(\theta_{n,h}=\theta_0+h/\sqrt n\) and
\[
 \mathcal L_n^{\mathrm{DA}}(h)
 :=n\{\widehat Q_{\mathrm{DA}}(\theta_{n,h})-\widehat Q_{\mathrm{DA}}(\theta_0)\}.
\]
Recall that $D_{n,h}(\cdot;\widehat{F})$ in Assumption \ref{Assump:PlugLocal}. In addition, $b_n=N/(n+N)$ and $r_n=n/N$.
\begin{lemma}[Local Quadratic Expansion]\label{lemma:Expansion1}
Under Assumptions \ref{Assump:first-step-L2} to \ref{Assump:Moments}, for every fixed \(M<\infty\), we have 
\begin{equation}
 \sup_{\|h\|\le M}
 \left|\mathcal L_n^{\mathrm{DA}}(h)-h^\top S_n-\frac12h^\top Hh\right|=\op(1),
 \label{eq:DA-local-quadratic}
\end{equation}
where
\begin{equation}
 S_n
 :=\G_n\{s_0-b_n\varsigma_0\}+b_n\sqrt{r_n}\,\G_N\varsigma_0+\op(1).
 \label{eq:Zn-DA}
\end{equation}
As a result, \(S_n=\Op(1)\).
\end{lemma}

\begin{proof}
We start with the following decomposition:
\begin{align}
&n\{\widehat Q_{\mathrm{DA}}(\theta_{n,h})-\widehat Q_{\mathrm{DA}}(\theta_0)\}\notag \\
=&n\left\{ a_n\Pn\ell_{\theta_{n,h}}+b_n\QN\mu_{\theta_{n,h}}(\,\cdot\,;\widehat F)-\left[ a_n\Pn\ell_{\theta_0}+b_n\QN\mu_{\theta_0}(\,\cdot\,;\widehat F) \right] \right\} \notag\\
=&na_n\Pn\left(\ell_{\theta_{n,h}}-\ell_{\theta_0}\right)+nb_n\QN\left(\mu_{\theta_{n,h}}(\,\cdot\,;F_0)-\mu_{\theta_0}(\,\cdot\,;F_0)\right)+nb_n\QN D_{n,h}(\cdot;\widehat{F})\notag\\
=&na_nP_0\left(\ell_{\theta_{n,h}}-\ell_{\theta_0}\right)+nb_nP_V\left(\mu_{\theta_{n,h}}(\,\cdot\,;F_0)-\mu_{\theta_0}(\,\cdot\,;F_0)\right)\label{eq:LocalQuadraticPL} \\
+&na_n(\Pn-P_0)\left(\ell_{\theta_{n,h}}-\ell_{\theta_0}\right)+nb_n(\QN-P_V)\left(\mu_{\theta_{n,h}}(\,\cdot\,;F_0)-\mu_{\theta_0}(\,\cdot\,;F_0)\right) \label{eq:LocalQuadraticEP}\\
+&nb_n\QN D_{n,h}(\cdot;\widehat{F})\label{eq:LocalQuadraticLF}.
\end{align}
The first two terms in the last equality (\ref{eq:LocalQuadraticPL}) involve the criterion function at the population level. Note that we have $a_nP_0\ell_\theta+b_nP_V\mu_\theta(\cdot;F_0)=Q(\theta)$. Assumption \ref{Assump:Smooth} yields uniformly over bounded \(h\) that,
\begin{equation}\label{eqn:PopulationQuadratic}
n\{Q(\theta_{n,h})-Q(\theta_0)\}=\frac12h^\top Hh+o(1),
\end{equation}
which contributes to the quadratic term in (\ref{eq:DA-local-quadratic}).

When it comes to the first term in (\ref{eq:LocalQuadraticEP}), we have
\begin{align}
 n a_n(\Pn-P_0)\{\ell_{\theta_{n,h}}-\ell_{\theta_0}\}
 =a_n h^\top\G_ns_0+\op(1), \label{eqn:LocalFluct1}
\end{align}
by Equation \eqref{eq:SE-labeled} in Assumption \ref{Assump:SE}. Similarly, the second term in (\ref{eq:LocalQuadraticEP}) admits the following expansion:
\begin{align}
 &n b_n(\QN-P_V)\{\mu_{\theta_{n,h}}(\cdot;F_0)-\mu_{\theta_0}(\cdot;F_0)\} \nonumber\\
 =&\sqrt{n}b_n\sqrt{\frac{n}{N}}\sqrt{N}(\QN-P_V)\{\mu_{\theta_{n,h}}(\cdot;F_0)-\mu_{\theta_0}(\cdot;F_0)\}\nonumber \\
 =&b_n\sqrt{r_n}\,h^\top\G_N\varsigma_0+\op(1), \label{eqn:LocalFluct2}
\end{align}
by Equation \eqref{eq:SE-unlabeled} in Assumption \ref{Assump:SE}.
 
It remains to account for estimation error of the conditional distribution stemmed from the last term (\ref{eq:LocalQuadraticLF}). By Assumption \ref{Assump:PlugLocal},
\[
 n b_n\QN D_{n,h}(\cdot;\widehat{F})
 =b_n h^\top\sqrt n\,\QN\Delta_n(\cdot;\widehat{F})+\op(1).
\]
Decompose
\[
 \sqrt n\,\QN\Delta_n(\cdot;\widehat{F})
 =\sqrt n\,P_V\Delta_n(\cdot;\widehat{F})+\sqrt n(\QN-P_V)\Delta_n(\cdot;\widehat{F}).
\]
The first term equals to \(\G_n(s_0-\varsigma_0)+\op(1)\) by \eqref{eq:first-step-AL}. Conditional on the labeled sample, we proceed with
\[
 \Var\!\left(\sqrt n(\QN-P_V)\Delta_n(\cdot;\widehat{F})\mid\{Z_i\}_{i=1}^n\right)
 \le r_n\|\Delta_n(\cdot;\widehat{F})\|_{P_V,2}^2=\op(1).
\]
Hence, $\sqrt n(\QN-P_V)\Delta_n(\cdot;\widehat{F})=o_p(1)$. Therefore, we have
\begin{equation}
 n b_n\QN D_{n,h}(\cdot;\widehat{F})=b_n\G_nh^\top(s_0-\varsigma_0)+\op(1). \label{eqn:LocalFluct3}
\end{equation}
Combining the expansions in Equations (\ref{eqn:LocalFluct1}), (\ref{eqn:LocalFluct2}), and (\ref{eqn:LocalFluct3}), three leading linear terms add up to
\begin{align*}
a_n\G_ns_0+b_n\G_n(s_0-\varsigma_0)+b_n\sqrt{r_n}\G_N\varsigma_0
=\G_n(s_0-b_n\varsigma_0)+b_n\sqrt{r_n}\G_N\varsigma_0.
\end{align*}
Together with the quadratic part (\ref{eqn:PopulationQuadratic}), this proves \eqref{eq:DA-local-quadratic}. The conclusion \(S_n=\Op(1)\) is deduced from the moment condition in Assumption \ref{Assump:Moments} and Lyapunov's CLT.
\end{proof}
 
\begin{proof}[Proof of Theorem \ref{thm:DA-AN}]
Let \(\widehat h_n=\sqrt n(\thetahat-\theta_0)\).  We first show that our estimator converges at the root-n rate.  Recall that in Assumption \ref{Assump:Smooth} we have \(\lambda_{\min}(H)>0\).  Since \(S_n=\Op(1)\), for any \(\varepsilon>0\) there is \(C<\infty\) such that \(\Pr(\|S_n\|\le C)>1-\varepsilon\) for all large \(n\).  Choose \(M>4C/\lambda_{\min}(H)\).  By Lemma \ref{lemma:Expansion1}, uniformly on \(\|h\|=M\),
\[
 \mathcal L_n^{\mathrm{DA}}(h)
 \ge -MC+\frac12\lambda_{\min}(H)M^2+o_\Pp(1)>0
 =\mathcal L_n^{\mathrm{DA}}(0)
\]
with probability at least \(1-2\varepsilon\) for a large enough \(n\).  Convexity of $\mathcal L_n^{\mathrm{DA}}(h)$ with respect to $h$ implies that $\Vert \widehat h_n\Vert\leq M$. Hence, \(\widehat h_n=\Op(1)\) follows.

Define \(h_n^*:=-H^{-1}S_n\), which is also \(\Op(1)\) by Lemma \ref{lemma:Expansion1}. We apply the expansion in Lemma \ref{lemma:Expansion1} twice by plugging in \(\widehat h_n\) and \(h_n^*\), respectively: 
\begin{align*}
    \mathcal L_n^{\mathrm{DA}}(\widehat h_n)&=\widehat h_n^\top S_n+\frac12\widehat h_n^\top H\widehat h_n+o_p(1)=-\widehat h_n^\top Hh_n^*+\frac12\widehat h_n^\top H\widehat h_n+o_p(1),\\
    \mathcal L_n^{\mathrm{DA}}(h_n^*)&=h_n^{*\top} S_n+\frac12\widehat h_n^{*\top} H\widehat h^*_n+o_p(1)=-\frac12\widehat h_n^\top H\widehat h_n+o_p(1),
\end{align*}
cf. Theorem 5.23 in \cite{vdVaart1998AsympStat}. Take the difference of the above two equations and note that $\mathcal L_n^{\mathrm{DA}}(\widehat h_n)\leq \mathcal L_n^{\mathrm{DA}}(h_n^*)$. Then we have
\[
 \frac12(\widehat h_n-h_n^*)^\top H(\widehat h_n-h_n^*)\leq\op(1).
\]
Positive definiteness of \(H\) in Assumption \ref{Assump:Smooth} yields that \(\widehat h_n-h_n^*=\op(1)\). Substituting the equation \eqref{eq:Zn-DA} for $S_n$, and using \(b_n\to b=(1+r)^{-1}\) and \(r_n\to r\), proves \eqref{eq:DA-linear-representation}.
The labeled and unlabeled empirical processes are independent in Assumption \ref{Assump:Sampling}. The multivariate CLT therefore gives covariance $ \Omega_r=\Var(s_0-b\varsigma_0)+b^2r\Var(\varsigma_0)$.
\end{proof}

\begin{proof}[Proof of Theorem \ref{thm:DDA-AN}]
As we have assumed the root-$n$ consistency, the crux is to establish the local quadratic expansion for 
\[
\mathcal L_n^{\mathrm{DDA}}(h)
:=n\{\widehat Q_{\mathrm{DDA}}(\theta_{n,h})-\widehat Q_{\mathrm{DDA}}(\theta_0)\}.
\]
We first show that for every fixed \(M<\infty\), we have
\begin{equation}
	\sup_{\|h\|\le M}
	\left|\mathcal L_n^{\mathrm{DDA}}(h)-h^\top S_n-\frac12h^\top Hh\right|=\op(1),
	\label{eq:DDA-local-quadratic}
\end{equation}	
where $S_n$ is the same as given in Equation \ref{eq:Zn-DA}.

To appreciate the role of the debiasing step, note that the following equality holds 
\[
b_n(P_V-P_0)\{\mu_{\theta_{n,h}}(\cdot;F)-\mu_{\theta_0}(\cdot;F)\}=0,
\]
for either $F=F_0$ or $F=\widehat F$. We proceed with the decomposition:
\begin{align*}
	&n\{\widehat Q_{\mathrm{DDA}}(\theta_{n,h})-\widehat Q_{\mathrm{DDA}}(\theta_0)\}\\
	=&n\Pn\left(\ell_{\theta_{n,h}}-\ell_{\theta_0}\right)+nb_n(\QN-\Pn)\left(\mu_{\theta_{n,h}}(\,\cdot\,;F_0)-\mu_{\theta_0}(\,\cdot\,;F_0)\right)+nb_n(\QN-\Pn) D_n(h,\cdot)\\
	=&nP_0\left(\ell_{\theta_{n,h}}-\ell_{\theta_0}\right)+n(\Pn-P_0)\left(\ell_{\theta_{n,h}}-\ell_{\theta_0}\right)\\
    +&nb_n(\QN-P_V)\left(\mu_{\theta_{n,h}}(\,\cdot\,;F_0)-\mu_{\theta_0}(\,\cdot\,;F_0)\right)-nb_n(\Pn-P_V)\left(\mu_{\theta_{n,h}}(\,\cdot\,;F_0)-\mu_{\theta_0}(\,\cdot\,;F_0)\right)\\
    +& nb_n(\QN-\Pn) D_{n,h}(\cdot;\widehat{F}).
\end{align*}
The first term on the right hand side of the last equality contributes to the quadratic term. Assumption \ref{Assump:SE} gives the linearization for the next three empirical process terms: \(\G_ns_0\), \(b_n\sqrt{r_n}\G_N\varsigma_0\), and \(-b_n\G_n \varsigma_0\), with asymptotically smaller order terms uniformly over bounded \(h\). Their sum is exactly \eqref{eq:Zn-DA}. The last term $ nb_n(\QN-\Pn)D_{n,h}(\cdot;\widehat{F})$ is asymptotically negligible by Assumption \ref{Assump:DDA-local}.

Given the quadratic expansion in \eqref{eq:DDA-local-quadratic}, we apply completion-of-the-square argument used in the proof of Theorem \ref{thm:DA-AN}, which yields
\[
 \sqrt n(\thetabc-\theta_0)=-H^{-1}S_n+\op(1).
\]
Equation \eqref{eq:DDA-linear-representation} follows from \eqref{eq:Zn-DA}, and the same CLT delivers the desired weak convergence to the desired normal limit \(N(0,\Sigma_r)\).
\end{proof}

\begin{proof}[Proof of Theorem~\ref{Thm:Bootstrap}]
Throughout the proof, let $\E^*$ denote expectation conditional on the original observations. Define
\begin{equation}
\widehat\Omega={}
 \frac1n\sum_{i=1}^n
 (\widehat v_i-\overline{\widehat v})
 (\widehat v_i-\overline{\widehat v})^\top
 +\frac{b_n^2r_n}{N}\sum_{j=n+1}^{n+N}
 (\widehat\varsigma_j-\overline{\widehat\varsigma}_U)
 (\widehat\varsigma_j-\overline{\widehat\varsigma}_U)^\top.
\label{eq:bootstrap-covariance}
\end{equation}
We consider an oracle statistic evaluated at the truth $(\theta_0,F_0)$ with multiplier bootstrap weights as follows.
\begin{footnotesize}
\begin{align*}
T_{n,0}^*:=-H^{-1}\Bigg[&
 \frac1{\sqrt n}\sum_{i=1}^n\xi_i
 \left\{s_0(Z_i)-b_n\varsigma_0(V_i)
              -\Pn(s_0-b_n\varsigma_0)\right\}+\frac{b_n\sqrt{r_n}}{\sqrt N}
 \sum_{j=n+1}^{n+N}\zeta_j
 \{\varsigma_0(V_j)-\QN\varsigma_0\}
 \Bigg].
\end{align*}
\end{footnotesize}
We proceed with
\begin{align*}
&\E^*\|T_n^*-T_{n,0}^*\|^2
\leq{} 2\|\widehat H^{-1}-H^{-1}\|_{\mathrm{op}}^2
       \operatorname{tr}(\widehat\Omega)\\
+&2\|H^{-1}\|_{\mathrm{op}}^2\Bigg[
 \frac1n\sum_{i=1}^n
  \bigl\|\widehat v_i-
       \{s_0(Z_i)-b_n\varsigma_0(V_i)\}\bigr\|^2
 +b_n^2r_n\frac1N\sum_{j=n+1}^{n+N}
        \|\widehat\varsigma_j-\varsigma_0(V_j)\|^2\Bigg]\\
 =&o_{\Pp}(1),
\end{align*}
The last equality follows from our Assumption \ref{ass:bootstrap-feasibility} (i) and (ii). Then, for every $f\in\operatorname{BL}_1(\mathbb R^p)$, we have
\[
 \left|\E^*f(T_n^*)-\E^*f(T_{n,0}^*)\right|
 \leq\E^*\|T_n^*-T_{n,0}^*\|
 \leq\{\E^*\|T_n^*-T_{n,0}^*\|^2\}^{1/2}
 =o_{\Pp}(1).
\]

Next, we prove the conditional weak convergence of the oracle statistic $T_{n,0}^*$, which has the same weak limit as $\sqrt n(\widehat\theta-\theta_0)$. We check the conditional version of the Lindeberg condition. For this purpose, we express 
\[
 T_{n,0}^*=\sum_{i=1}^n\xi_i\psi_{n,i}
             +\sum_{j=n+1}^{n+N}\zeta_j\psi_{n,j},
\]
where
\begin{equation}
\psi_{n,k}:=
\begin{cases}
-\dfrac{H^{-1}}{\sqrt n}
 \bigl\{s_0(Z_k)-b_n\varsigma_0(V_k)
             -\Pn(s_0-b_n\varsigma_0)\bigr\},
 &1\leq k\leq n,\\[2mm]
-\dfrac{b_n\sqrt{r_n}\,H^{-1}}{\sqrt N}
 \bigl\{\varsigma_0(V_k)-\QN\varsigma_0\bigr\},
 &n+1\leq k\leq n+N.
\end{cases}
\label{eq:bootstrap-oracle-coefficients}
\end{equation}
Under our Assumption \ref{ass:bootstrap-feasibility}, it can be shown that \begin{equation}
 \left\|\sum_{k=1}^{n+N}\psi_{n,k}\psi_{n,k}^\top
          -\Sigma_r\right\|_{\mathrm{op}}\to_p0,
\label{eq:bootstrap-oracle-variance-limit}
\end{equation} and $\sum_{k=1}^{n+N}\|\psi_{n,k}\|^2=O_p(1)$. Then, for every $\varepsilon>0$, we have
\[
 n\Pp\{\|s_0(Z)\|>\varepsilon\sqrt n\}
 \leq\varepsilon^{-2}
 \E\!\left[\|s_0(Z)\|^2
        \mathbf 1\{\|s_0(Z)\|>\varepsilon\sqrt n\}\right]\to0.
\]
A union bound therefore yields
$\max_{i\leq n}\|s_0(Z_i)\|/\sqrt n\to_p0$.
The same argument applies to $\varsigma_0$ in both labeled and unlabeled samples. It follows that
\begin{equation}
 \max_{1\leq k\leq n+N}\|\psi_{n,k}\|\to_p0.
\label{eq:bootstrap-oracle-max-limit}
\end{equation}
Fix
$\varepsilon>0$ and $M>0$. On the event
$\max_k\|\psi_{n,k}\|\leq\varepsilon/M$, the sum
\[
 \sum_{i=1}^n\E^*\!\left[
    \|\xi_i\psi_{n,i}\|^2
    \mathbf 1\{\|\xi_i\psi_{n,i}\|>\varepsilon\}\right]
 +\sum_{j=n+1}^{n+N}\E^*\!\left[
    \|\zeta_j\psi_{n,j}\|^2
    \mathbf 1\{\|\zeta_j\psi_{n,j}\|>\varepsilon\}\right]
\]
is bounded above by
\begin{equation}
 \left(\sum_{k=1}^{n+N}\|\psi_{n,k}\|^2\right)
 \left\{
  \E[\xi_1^2\mathbf 1\{|\xi_1|>M\}]
  +\E[\zeta_{n+1}^2\mathbf 1\{|\zeta_{n+1}|>M\}]
 \right\}.
\label{eq:bootstrap-oracle-lindeberg-bound}
\end{equation}
Both expectations in the curly bracket vanish as $M\to\infty$ under our Assumption \ref{ass:bootstrap-feasibility} (iii).

Finally, one can argue as in \citet[proof of Theorem~3.2, Step~VI]{LinHan2026BootstrapDML} that every subsequence has a further subsequence along which
\eqref{eq:bootstrap-oracle-variance-limit} and
\eqref{eq:bootstrap-oracle-max-limit} hold almost surely. For almost every
realization of the observations along that subsequence, the coefficients
are deterministic, and the bound~\eqref{eq:bootstrap-oracle-lindeberg-bound}
verifies the Lindeberg condition for the independent multiplier array. Thereafter, we have
\begin{equation*}
    \sup_{f\in\operatorname{BL}_1(\mathbb R^p)}
 \left|\E^*f(T_{n,0}^*)-
   \E f\{\sqrt n(\widehat\theta-\theta_0)\}\right|\to_p 0,
\end{equation*}
which completes the proof.
\end{proof}

\section{Asymptotic Linear Representation of DA}\label{app:small-bias}

\subsection{Linear Representation for Local Linear Estimators}
\label{subsec:small-bias-kernel-series}
In this part, we provide the formal proof of the linear representation for the local linear kernel smoothing method for the conditional distribution estimation, which is used in our Monte Carlo simulation. We provide primitive conditions on the underlying distribution, score function, kernel function, and choices of bandwidth. 

We first provide a heuristic argument using the local constant smoothing approach which is more transparent. We temporarily bypass the technical issues to convey the main idea. A rigorous argument can be adapted from the proof of (A.1) in \cite{WangChen2009EmpiricalLikelihoodMissing}. For a generic kernel function satisfying $\int K(u)\,du=1$, let $K_h(u)=h^{-d_V}K(u/h)$. The local constant estimator for the conditional distribution is
\begin{equation}
 \widehat F^{LC}_{h}(y\mid v)
 =\frac{\sum_{i=1}^n K_h(V_i-v)\mathbf1\{Y_i\le y\}}
        {\sum_{i=1}^n K_h(V_i-v)}.
 \label{eq:LC-heuristic-estimator}
\end{equation}
Modulo the bias term, we have the uniform linear representation as follows:
\begin{equation}
 \widehat F^{LC}_{h}(y\mid v)-F_0(y\mid v)
 \simeq\frac{1}{nf_V(v)}\sum_{i=1}^n K_h(V_i-v)
       \{\mathbf1\{Y_i\le y\}-F_0(y\mid V_i)\},
 \label{eq:LC-heuristic-CDF}
\end{equation}
where $f_V$ stands for the marginal density of $V$. Therefore, we get 
\begin{equation}
\begin{aligned}
 P_V\Delta_n(\cdot;\widehat F^{LC}_{h})
 &:=\int_{\mathcal V}\!\int s_0(x,y)\,
           d\{\widehat F^{LC}_{h}(y\mid v)-F_0(y\mid v)\}\,dP_V(v)\\
 &\simeq\frac1n\sum_{i=1}^n\int_{\mathcal V}K_h(V_i-v)
       \left[\int s_0(x,y)\,
       d\{\mathbf1\{Y_i\le y\}-F_0(y\mid V_i)\}\right]dv.
\end{aligned}
 \label{eq:LC-heuristic-score-integral}
\end{equation}
For each summand, one can apply change of variable and explore the smoothness of the score function over $x$ to get
\begin{equation}
\begin{aligned}
 &\int_{\mathcal V}K_h(V_i-v)
       \left[\int s_0(x,y)\,
       d\{\mathbf1\{Y_i\le y\}-F_0(y\mid V_i)\}\right]dv\\
  \simeq&\{s_0(X_i,Y_i)-\int s_0(X_i,y)\,dF_0(y\mid V_i)\}\underbrace{\int K(u)\,du}_{=1},
\end{aligned}
 \label{eq:LC-heuristic-localization}
\end{equation}
which leads to the desired formula in the asymptotic linear representation. 

Now we consider the local linear estimator. The covariate kernel function is taken to be the product kernel of each coordinates, i.e., $K(u)=\prod_{j=1}^{d_V}K_j(u_j)$. Let $r(u)=(1,u^\top)^\top$ and let $e_0=(1,0,\ldots,0)^\top$ select the intercept. Define
\begin{align}
 \widehat\Gamma_h(v)
 &=\frac1n\sum_{i=1}^n K_h(V_i-v)
       r((V_i-v)/h)r((V_i-v)/h)^\top,\nonumber\\
 \Gamma_h(v)&=\E\widehat\Gamma_h(v)=\int_{\{u\in[-1,1]^{d_V}:v+hu\in\mathcal V\}}
       K(u)r(u)r(u)^\top f_V(v+hu)\,du.,\label{eq:kernel-LL-matrices}\\
 \widehat F^{\mathrm{LL}}_{h}(y\mid v)
 &=e_0^\top\widehat\Gamma_h(v)^{-1}
       \frac1n\sum_{i=1}^n K_h(V_i-v)r((V_i-v)/h)\mathbb I\{Y_i\leq y\}.
 \label{eq:kernel-LL-CDF}
\end{align}
Furthermore, we set
\[
 \widehat W_{i,h}(v)=K_h(V_i-v)e_0^\top\widehat\Gamma_h(v)^{-1}r((V_i-v)/h),
\]
and its population counterpart as 
\[
 W_h(v,w)=K_h(w-v)e_0^\top\Gamma_h(v)^{-1}r((w-v)/h).
\]
For every linear function $L(\cdot)$, the local linear smoothing has the following reproducing property
\begin{equation}
 \frac1n\sum_i\widehat W_{i,h}(v)L(V_i)=L(v).
 \label{eq:kernel-LL-reproduction}
\end{equation}
See Lemma 11 in \cite{FanGuerre2016Multivariate}.

We impose the following primitive conditions, adapted from \cite{FanGuerre2016Multivariate} and \cite{Xie2023Uniform}. We do not claim these are the weakest possible conditions. For instance, the bandwidth sequence requires the standard under-smoothing. One can also adapt the twicing kernel approach of \cite{NeweyHsiehRobins2004Twicing} to relax this condition. The local linear estimates may not guarantee the positive weights in finite sample, and one may resort to the modification by \cite{HallWolffYao1999Conditional}. In addition, one can also smooth over the outcome $Y$ as in \cite{Xie2023Uniform}. 
\begin{assumption}\label{ass:kernel-benchmark}
\begin{assumpenum}
\item\label{cond:kernel-design}
The labeled $Z_i=(Y_i,V_i)$ are i.i.d. from $P_0$. The support $\mathcal V\subset\R^{d_V}$ is compact, convex, and has nonempty interior. Its density $f_V$ satisfies $0<c\leq f_V(v)\leq C<\infty$ on $\mathcal V$ and has a continuously differentiable extension to a neighborhood of $\mathcal V$. There are constants $\lambda_0,\lambda_1>0$ such that, for every $v\in\mathcal V$ and $0<\epsilon\leq\lambda_0$, some ball of radius $\lambda_1\epsilon$ lies in $B(v,\epsilon)\cap\mathcal V$. The boundary $\{v\in\mathcal V:\operatorname{dist}(v,\partial\mathcal V)\leq h\}$ has Lebesgue measure $O(h)$. In addition, $f_V$ has bounded second derivatives on this set.
\item\label{cond:kernel-functions}
The covariate kernel function is $K(u)=\prod_{j=1}^{d_V}K_j(u_j)$, where each $K_j$ is a bounded, symmetric, nonnegative, continuous density supported on $[-1,1]$, positive on $(-1,1)$, and of bounded variation. Set $K_h(u)=h^{-d_V}K(u/h)$. 
\item\label{cond:kernel-smoothness}
$F_0(y\mid v)$ has an extension in $v$ with bounded first and second derivatives, uniformly in $y$, such that its first-order Taylor remainder is bounded by $C\|w-v\|^2$ uniformly in $y$ for nearby $v,w$ in the support. 
\item\label{cond:kernel-score}
$Y$ is supported on a fixed compact interval. For some $\alpha>0$,
\[
 \int\|s_0(x,y)-s_0(x',y)\|^2\,dF_0(y\mid w)
 \leq C\|v-w\|^{2\alpha},\qquad v,w\in\mathcal V\text{ sufficiently close}.
\]
\end{assumpenum}
\end{assumption}
\begin{proposition}\label{prop:kernel-LL-AL}
Let Assumption~\ref{ass:kernel-benchmark} hold. Suppose
\begin{equation}
 h\to0,\qquad \frac{nh^{d_V}}{\log n}\to\infty,\qquad
 \sqrt nh^2\to0.
 \label{eq:kernel-LL-bandwidth}
\end{equation}
Then, we have
\begin{equation}
 \sqrt n\{\Psi(\widehat F^{\mathrm{LL}}_{h})-\Psi(F_0)\}
 =\G_n\{s_0-\varsigma_0\}+\op(1).
 \label{eq:kernel-LL-AL}
\end{equation}
\end{proposition}
\begin{proof}
We start with the following decomposition:
\begin{equation}
\begin{aligned}
 &\sqrt n\{\Psi(\widehat F^{\mathrm{LL}}_{h})-\Psi(F_0)\}
       -\G_n(s_0-\varsigma_0)\\
 ={}&\underbrace{\sqrt n\int_{\mathcal V}\!\int s_0(x,y)\,
       d\left\{\frac1n\sum_i\widehat W_{i,h}(v)F_0(y\mid V_i)
                         -F_0(y\mid v)\right\}dP_V(v)}_{\textcircled{1}}\\
 &+\underbrace{\frac1{\sqrt n}\sum_i\int_{\mathcal V}
       \{\widehat W_{i,h}(v)-W_h(v,V_i)\}
       \left[s_0(x,Y_i)-\int s_0(x,y)\,dF_0(y\mid V_i)\right]dP_V(v)}_{\textcircled{2}}\\
 &+\underbrace{\frac1{\sqrt n}\sum_i\int_{\mathcal V}W_h(v,V_i)
       \left[
       s_0(x,Y_i)-s_0(X_i,Y_i)
       -\int\{s_0(x,y)-s_0(X_i,y)\}\,dF_0(y\mid V_i)
       \right]dP_V(v)}_{\textcircled{3}}\\
 &+\underbrace{\frac1{\sqrt n}\sum_i\{s_0(X_i,Y_i)-\varsigma_0(V_i)\}
       \left\{\int_{\mathcal V}W_h(v,V_i)\,dP_V(v)-1\right\}}_{\textcircled{4}},
\end{aligned}
\label{eq:kernel-LL-four-term-decomposition}
\end{equation}
where $\varsigma_0(V_i)=\int s_0(X_i,y)\,dF_0(y\mid V_i)$.

\emph{Step 1.} We bound the first term $\textcircled{1}$, which is the approximation bias.
Conditional on the covariate, $\widehat F^{\mathrm{LL}}_{h}$ has mean $n^{-1}\sum_i\widehat W_{i,h}(v)F_{0}(y\mid V_i)$. The standard Taylor expansion gives us that
\[
 F_{0}(y\mid V_i)=F_{0}(y\mid v)
 +\nabla_v F_{0}(y\mid v)^\top(V_i-v)+R_i(y,v),
 \qquad |R_i(y,v)|\le C\|V_i-v\|^2.
\]
Referring to \eqref{eq:kernel-LL-reproduction}, we have $n^{-1}\sum_i\widehat W_{i,h}(v)=1$ and $n^{-1}\sum_i\widehat W_{i,h}(v)(V_i-v)=0$. Subtracting $F_{0}(y\mid v)$ from the conditional mean therefore leaves exactly $n^{-1}\sum_i\widehat W_{i,h}(v)R_i(y,v)$. Consequently, we have
\[
 \sup_{y,v}\left|\E[\widehat F^{\mathrm{LL}}_{h}(y\mid v)\mid V_1,\ldots,V_n]-F_0(y\mid v)\right|
 \leq Ch^2.
\]

\emph{Step 2.} We replace the sample weighting matrix by its population counterpart in each centered term. Assumption \ref{ass:kernel-benchmark} (i) implies that $\widehat\Gamma_h(v)$ is invertiable with probability one. The maximal inequality in \citet[Lemma 2]{Xie2023Uniform} therefore gives 
\[
 \sup_v\|\widehat\Gamma_h(v)^{-1}-\Gamma_h(v)^{-1}\|_{op}
 =\Op\!\left(\sqrt{\frac{\log n}{nh^{d_V}}}
                   +\frac{\log n}{nh^{d_V}}\right)=\op(1),
\]
which leads to
\[
 \int_{\mathcal V}|\widehat W_{i,h}(v)-W_h(v,V_i)|\,dP_V(v)
 \le C\sup_v\|\widehat\Gamma_h(v)^{-1}-\Gamma_h(v)^{-1}\|_{op}.
\]
Hence, we proceed with
\begin{align}
 &\E\!\left[
 \left\|\frac1{\sqrt n}\sum_i\int_{\mathcal V}
       \{\widehat W_{i,h}(v)-W_h(v,V_i)\}
       \left[s_0(x,Y_i)-\int s_0(x,y)\,dF_0(y\mid V_i)\right]dP_V(v)
 \right\|^2\,\middle|\,(V_i)_{i=1}^n\right]\nonumber\\
 &\quad\le\frac Cn\sum_i
       \left\{\int_{\mathcal V}
       |\widehat W_{i,h}(v)-W_h(v,V_i)|\,dP_V(v)\right\}^2
 \le C\sup_v\|\widehat\Gamma_h(v)^{-1}-\Gamma_h(v)^{-1}\|^2_{op}
 =\op(1),
 \label{eq:kernel-LL-matrix-replacement-variance}
\end{align}
which proves the negligibility of the term $\textcircled{2}$ in \eqref{eq:kernel-LL-four-term-decomposition}.

\emph{Step 3.} When it comes to the term $\textcircled{3}$, we make use of the smoothness of the score function in Assumption \ref{ass:kernel-benchmark}:
\begin{align*}
 &\E\!\left[
 \left\|s_0(x,Y_i)-s_0(X_i,Y_i)
       -\int\{s_0(x,y)-s_0(X_i,y)\}\,dF_0(y\mid V_i)\right\|^2
 \,\middle|\,V_i\right]\\
 &\qquad\le\E[\|s_0(x,Y_i)-s_0(X_i,Y_i)\|^2\mid V_i]
 \le C\|v-V_i\|^{2\alpha}.
\end{align*}
This in turn provides the desired bound as follows:
\begin{align}
 &\E\left\|\frac1{\sqrt n}\sum_i\int_{\mathcal V}W_h(v,V_i)
 \left[s_0(x,Y_i)-s_0(X_i,Y_i)
-\int\{s_0(x,y)-s_0(X_i,y)\}\,dF_0(y\mid V_i)
 \right]dP_V(v)\right\|^2\nonumber\\
 &\qquad\le\frac{Ch^{2\alpha}}n\sum_i
       \E\left\{\int_{\mathcal V}|W_h(v,V_i)|\,dP_V(v)\right\}^2
 \le Ch^{2\alpha}\to0.
 \label{eq:kernel-LL-score-location-variance}
\end{align}
\emph{Step 4.} Referring to the last term $\textcircled{4}$, we have
\begin{equation}
 \int_{\mathcal V}W_h(v,w)\,dP_V(v)=\int_{\{u\in[-1,1]^{d_V}:w-hu\in\mathcal V\}}
 K(u)f_V(w-hu)e_0^\top\Gamma_h(w-hu)^{-1}r(u)\,du.
 \label{eq:kernel-LL-exact-integrated-weight}
\end{equation}
by the change of variable. We need to distinguish the interior versus boundary cases. For this purpose, we define
\[
 \mathcal V_h^\circ
 =\{w\in\mathcal V:\operatorname{dist}(w,\partial\mathcal V)>2\sqrt{d_V}h\},
 \qquad \mathcal B_h=\mathcal V\setminus\mathcal V_h^\circ.
\]
Under Assumption \ref{ass:kernel-benchmark} (i), it can be shown that 
\begin{equation}
 \sup_{w\in\mathcal V_h^\circ}
 \left|\int_{\mathcal V}W_h(v,w)\,dP_V(v)-1\right|
 \le Ch\int K(u)\,du=Ch,
 \label{eq:kernel-LL-interior-weight}
\end{equation}
and
\begin{equation}
 P_V(\mathcal B_h)
 \le\sup f_V\,\operatorname{Leb}
       \{w\in\mathcal V:\operatorname{dist}(w,\partial\mathcal V)
                                     \le2\sqrt{d_V}h\}
 \le Ch.
 \label{eq:kernel-LL-boundary-strip}
\end{equation}
For either set $\mathcal A=\mathcal V_h^\circ$ or $\mathcal B_h$, we have
\begin{align}
 &\E\left\|\frac1{\sqrt n}\sum_i
       \{s_0(X_i,Y_i)-\varsigma_0(V_i)\}
       \left\{\int_{\mathcal V}W_h(v,V_i)\,dP_V(v)-1\right\}
       \mathbf1\{V_i\in\mathcal A\}\right\|^2\nonumber\\
 &\quad=\E\!\left[
       \|s_0(X_i,Y_i)-\varsigma_0(V_i)\|^2
       \left|\int_{\mathcal V}W_h(v,V_i)\,dP_V(v)-1\right|^2
       \mathbf1\{V_i\in\mathcal A\}\right]\nonumber\\
 &\quad\le
 \begin{cases}
 Ch^2,&\mathcal A=\mathcal V_h^\circ,\\
 C P_V(\mathcal B_h)\le Ch,&\mathcal A=\mathcal B_h.
 \end{cases}
 \label{eq:kernel-LL-weight-normalization-variance}
\end{align}
In sum, all four terms in \eqref{eq:kernel-LL-four-term-decomposition} are negligible under \eqref{eq:kernel-LL-bandwidth}, proving \eqref{eq:kernel-LL-AL}.
\end{proof}

\subsection{Influence Function Perspective}
Here we present the calculation of the influence function for the linear functional $\Psi(F)$. Although this calculation does not automatically grant the stochastic expansion of an arbitrary $\widehat{F}$. Nevertheless, it ensures the form of the linear representation in Assumption \ref{ass:first-step-AL} is agnostic to the first-step estimation \citep{Newey1994AsymptoticVariance}. The particular functional involved is related to the example in Section 7 of \cite{severini2001SimplifiedApproachEfficiencyBounds} in which they study the information bound for a linear function of the conditional expectation weighted by a given weighting function. 

Recall that $Z=(Y,V)$, $V=(X,\widehat Y)$, and $v=(x,\widehat y)$. Let $P_0(dy,dv)=F_0(dy\mid v)P_V(dv)$ be the joint law of the labeled-data and write $s_0(x,y)=s_{\theta_0}(x,y)$ as the score function at the truth. Define the linear functional as
\begin{equation}
 \Psi(F)=\int\!\int s_0(x,y)\,dF(y\mid v)\,dP_V(v).
 \label{eq:IF-fixed-reference-functional}
\end{equation}
For this calculation, both $s_0$ and the marginal law $P_V$ are held fixed. We keep using the short-hand notation $\varsigma_0(V)=\int s_0(X,y)\,dF_0(y\mid V)$.
 
Let $L_2(P_0)$ denote the real Hilbert space with inner product $\langle g,h\rangle=\E[gh]$, and let $L_2^0(P_0)$ be the subspace containing zero mean functions. For the local conditional submodels $F_\epsilon$ (that are dominated) through $F_0$, define the likelihood ratio
\[
 r_\epsilon(y,v)=\frac{dF_\epsilon(\cdot\mid v)}{dF_0(\cdot\mid v)}(y).
\]
We use paths with score $S\in L_2(P_0)$ satisfying
\begin{equation}
 \left\|\frac{r_\epsilon-1}{\epsilon}-S\right\|_{P_0,2}\longrightarrow0.
 \label{eq:IF-path-regularity}
\end{equation}
\begin{lemma}[Canonical Gradient]\label{lemma:IF}
Suppose $\E\|s_0(X,Y)\|^2<\infty$. The tangent space is
\begin{equation}
 \mathcal T_{Y\mid V}
 =\{S\in L_2^0(P_0):\E[S\mid V]=0\}.
 \label{eq:IF-conditional-tangent}
\end{equation}
The functional in \eqref{eq:IF-fixed-reference-functional} has continuous pathwise derivative as
\begin{equation}
 \left.\frac{d}{d\epsilon}\Psi(F_\epsilon)\right|_{\epsilon=0}
 =\E[\phi_0(Z)S(Z)],
 ~~~\text{where}~~
 \phi_0(Z)=s_0(X,Y)-\varsigma_0(V).
 \label{eq:IF-canonical-gradient}
\end{equation}
\end{lemma}
\begin{proof}
    We first show the tangent space is given by \eqref{eq:IF-conditional-tangent}. The definition of the conditional likelihood ratio gives $\E[r_\epsilon\mid V]=1$, which in turn implies $\E[S\mid V]=0$. Note that this defines a closed subspace of $L_2(P_0)$. Conversely, take any bounded function $S$ with $\E[S\mid V]=0$. For $|\epsilon|\|S\|_\infty<1$, set
\begin{equation}
 F_\epsilon(y\mid v)=\int_{-\infty}^y
          \{1+\epsilon S(t,v)\}\,dF_0(t\mid v).
 \label{eq:IF-conditional-measure-tilt}
\end{equation}
The integrand is positive and integrates to one conditionally on $V$, which produces a valid conditional distribution. In addition, $(r_\epsilon-1)/\epsilon=S$ exactly. The extension from the bounded $S$ to those in the $L_2$ space can be obtained via the standard denseness argument.

Now for any path satisfying \eqref{eq:IF-path-regularity}, we have
\[
 \frac{\Psi(F_\epsilon)-\Psi(F_0)}{\epsilon}
 =\E\!\left[s_0(X,Y)\frac{r_\epsilon(Z)-1}{\epsilon}\right].
\]
By the assumption that $\E\|s_0(X,Y)\|^2<\infty$ and the Cauchy--Schwarz inequality, we have
\[
 \left\|\frac{\Psi(F_\epsilon)-\Psi(F_0)}{\epsilon}
                     -\E[s_0(X,Y)S(Z)]\right\|
 \le\|s_0\|_{P_0,2}
       \left\|\frac{r_\epsilon-1}{\epsilon}-S\right\|_{P_0,2}
 \longrightarrow0.
\]
Moreover, the defining property of the tangent set gives us
\[
 \E[\varsigma_0(V)S(Z)]
 =\E[\varsigma_0(V)\E[S\mid V]]=0.
\]
Finally, subtracting the above zero term proves \eqref{eq:IF-canonical-gradient}.
\end{proof}

\section{Stochastic Equicontinuity of DDA}\label{Sec: DDA and DeepLearning}
In this part, we provide an in-depth analysis on the stochastic equicontinuity. We verify it by the empirical process theory \citep{vdVaartWellner2023EP} given the rate of convergence $\epsilon_n$ for the conditional distribution estimation. Assume that $\widehat{F}\in \mathcal{F}_n$ with probability one, for certain functional class $\mathcal{F}_n$ to be detailed in the sequel. Because of the consistency of the point estimation, we can focus on a fixed closed ball $B\subset\operatorname{int}(\Theta)$ centered at $\theta_0$. 

\begin{assumption}\label{ass:dda-dnn}
\begin{enumerate}[label=(\roman*),leftmargin=*]
\item The loss function satisfies
\begin{equation}
 \ell_{\theta+t}(x,y)-\ell_\theta(x,y)
 =\int_0^1t^\top s_{\theta+ut}(x,y)\,du.
 \label{eq:dda-dnn-segment}
\end{equation}
The measurable envelope $L(x,y)$, which dominates
$|\ell_{\theta_0}(x,y)|+\sup_{\theta\in B}\|s_\theta(x,y)\|$, satisfies$\int L(x,y)\,dF(y\mid v)\le C<\infty$ for some finite positive constant $C$.
\item The estimated conditional distribution $\widehat F\in \mathcal F_n$ with probability tending to one. Its rate of convergence by the following metric is
\[
 \sup_{\theta\in B}\|\varsigma_\theta(v;\widehat F)-\varsigma_\theta(v;F_0)\|_{P_V,2}=O_P(\epsilon_n),
\]
form some $\epsilon_n\to0$. 
\item Consider the following functional class $ \mathcal H_n=\{\varsigma_{\theta_j}(v;F)-\varsigma_{\theta_j}(v;F_0):
       \theta\in B,\ F\in\mathcal F_n,\ 1\le j\le p\}$. We assume its bracketing entropy number satisfies
\begin{equation}
 1+\log N_{[]}(\eta,\mathcal H_n,L_2(P_V))
 \le C S_n\log(A_n/\eta),\quad 0<\eta<1/2,
 \qquad S_n\ge1,\ A_n\ge e.
 \label{eq:dda-dnn-entropy}
\end{equation}
\item The rate of convergence and the upper bound for the entropy number satisfy the following:
\begin{equation}
 S_n\log(nA_n)=o(\sqrt n),\qquad S_n\log(nA_n)\epsilon_n^2\longrightarrow0,
 \label{eq:dda-dnn-rate-conditions}
\end{equation}
\end{enumerate}
\end{assumption}

\begin{proposition}\label{prop:dda-dnn}
Suppose Assumption~\ref{ass:dda-dnn} holds. Then, we have
\begin{align*}
    \sup_{\|h\|\le M}\sqrt{n}\left|(\QN-\Pn)h^\top\Delta_n(\cdot;\widehat{F})\right|=o_p(1).
\end{align*}
\end{proposition}

\begin{proof}[Proof of Proposition \ref{prop:dda-dnn}]
To slightly abuse the notation, we write $ \Delta_\theta(v;F)=\varsigma_\theta(v;F)-\varsigma_\theta(v;F_0)$ and denote its $j$-th coordinate as $\Delta_{\theta,j}(\cdot;F)$ for $j=1,\cdots,p$. The first condition in our assumption allows us to express
\begin{equation*}
    D_{n,h}(v;\widehat F)
 =\frac1{\sqrt n}\int_0^1h^\top
       \Delta_{\theta_0+uh/\sqrt n}(v;\widehat F)\,du.
\end{equation*}
Recall that two empirical measures have the same population marginal $P_V$. By subtracting and adding $P_Vh^\top\Delta_n(\cdot;\widehat{F})$, we have
\begin{align*}
 n(\mathbb Q_N-\mathbb P_n)D_{n,h}
 =\int_0^1h^\top\bigl\{
       \sqrt{n/N}\,\mathbb G_N\Delta_{\theta_0+uh/\sqrt n}
       -\mathbb G_n\Delta_{\theta_0+uh/\sqrt n}\bigr\}\,du.
\end{align*}
Taking the supremum over $\|h\|\le M$, it suffices to prove that 
\begin{equation*}
    \sup_{\theta\in B}|\mathbb G_n\Delta_{\theta,j}(\cdot;\widehat F)|
 =o_P(1),~~~
 \sqrt{n/N}\sup_{\theta\in B}
 |\mathbb G_N\Delta_{\theta,j}(\cdot;\widehat F)|
 =o_P(1).
\end{equation*}
Without loss of generality, we take $j$-th coordinate from the class $\mathcal H_n$ and we apply the maximal inequality from Theorem~2.14.2 in \cite{vdVaartWellner2023EP}. For $\rho>0$, restrict $\mathcal H_n$ to functions with norm in Assumption \ref{ass:dda-dnn} (ii) at most $\rho$. It has a constant envelope $B_*\ge1$, since $|\Delta_{\theta,j}|\le2\mathcal E$. Write
$H_n(\eta)=C S_n\log(A_n/\eta)$ for its entropy upper bound. We also introduce $c_n=S_n\log(nA_n)$ to simplify the notation. This gives
\begin{equation}
 \mathbb E\sup_{\|g\|_{P_V,2}\le\rho,\ g\in\mathcal H_n}|\mathbb G_n g|
 \le C\left\{\int_0^\delta\sqrt{H_n(\eta)}\,d\eta
                  +\rho\sqrt{H_n(\delta)}\right\},
 \quad \sqrt n\,\delta>B_*\sqrt{H_n(\delta)}.
 \label{eq:dda-dnn-maximal}
\end{equation}
Take $\delta=C_*B_*\sqrt{c_n/n}$, with $C_*$ sufficiently large. Then $\delta\to0$, $H_n(\delta)\le Cc_n$. To evaluate the integral, substitute $\eta=\delta t$ and use the elementary inequality that $\sqrt{a+b}\le\sqrt a+\sqrt b$:
\[
 \int_0^\delta\sqrt{S_n\log(A_n/\eta)}\,d\eta
 \le\delta\sqrt{S_n\log(A_n/\delta)}
   +\delta\sqrt{S_n}\int_0^1\sqrt{\log(1/t)}\,dt
 \le C\delta\sqrt{c_n}.
\]
Thus, we have
\begin{equation}
\mathbb E\sup_{\|g\|_2\le\rho,\ g\in\mathcal H_n}|\mathbb G_n g|
 \le C\{c_n/\sqrt n+\rho\sqrt{c_n}\},
 \label{eq:dda-dnn-local-bound}
\end{equation}
where the right hand side of the inequality converges to zero under the last condition in our Assumption \ref{ass:dda-dnn}. The proof for the unlabeled part follows mutatis mutandis. 
\end{proof}
Now we discuss more primitive conditions tailored to the deep learning methods. First, verification of the rate of convergence defined by the metric $\Vert \cdot\Vert_{P_V,2}$ can be translated into more familiar norms for the conditional distribution. Consider the following $L_2$-type metric
\[
 d_{L_2}(F,F_0)
   =\sqrt{P_V\int|F(y\mid V)-F_0(y\mid V)|^2\,dy},
\]
cf. Equation (4.2) in \cite{ChaiFanGu2026NDRE}. If the scores are absolutely continuous in $y$ and
$\sup_{\theta,x}\|\partial_y s_{\theta,j}(x,\cdot)\|_{L_2}\le C$ for any $j$, the Cauchy--Schwarz inequality yields
$\sup_{\theta\in B}\|\varsigma_\theta(v;\widehat F)-\varsigma_\theta(v;F_0)\|_{P_V,2}\le Cd_{L_2}(\widehat F,F_0)$. Thus, the convergence rate of \cite{ChaiFanGu2026NDRE} can be used to verify Part (ii) in Assumption \ref{ass:dda-dnn}\footnote{For other DNN-based distribution estimators, we refer readers to \cite{Liang2021How}, \cite{TangYang2023Minimax}, and \cite{PuchkinEtAl2024Rates}. }. 

When it comes to the bracketing entropy number, one can further assume that the score function satisfies
\begin{equation*}
    \sup_{y}\|s_\theta(x,y)-s_{\theta'}(x,y)\|
 \le \mathcal{E}(v)\|\theta-\theta'\|,
\end{equation*}
for some function $\mathcal{E}\in L_2(P_V)$. We also introduce the stronger metric for the conditional distribution functions:
\[
 d_{2,\infty}(F,G)
 =\operatorname*{ess\,sup}_{v\sim P_V}
       \left\{\int|F(y\mid v)-G(y\mid v)|^2\,dy\right\}^{1/2}.
\]
For each coordinate and generic pairs of $(\theta,F)$ and $(\theta^\prime,F^\prime)$, we have the following bound
\begin{equation}
 |\Delta_{\theta,j}(v;F)-\Delta_{\theta',j}(v;F^\prime)|
 \le C_1 d_{2,\infty}(F,F^\prime)
       +C_2\mathcal{E}(v)\|\theta-\theta'\|,
 \label{eq:dda-dnn-pointwise-transfer}
\end{equation}
some finite constant terms $C_1$ and $C_2$. This implies that 
\begin{equation}
 \log N_{[]}(\eta,\mathcal H_n,L_2(P_V))
 \le \log p+\log N(c\eta,\mathcal F_n,d_{2,\infty})
                       +p\log(C/\eta).
 \label{eq:dda-dnn-entropy-transfer}
\end{equation}


Therefore, the crux is to bound the proper entropy number of the deep learning architecture. For concreteness, we use the deep neural network (DNN) with ReLU activation function $\sigma(x)=\max\{x,0\}$ as the basis function as our illustration \citep{SchmidtHieber2020Nonparametric,KohlerLanger2021Rate}. We explore DNN to learn a function $\widehat{g}$ such that $\widehat{g}(V=v,U)$ follows the same conditional distribution of the scalar outcome $Y$ given $V=v$ for a bounded noise vector $U\in \mathbb{R}^{d_U}$. Let $L$ and $W$ be any positive integers. The deep neural network with depth $L$ and ambient width $W$ admits the following form:
\begin{equation*}	g(v,u)=\mathcal{T}_{L+1}\circ\bar{\sigma}\circ\mathcal{T}_L\circ\cdots\circ \mathcal{T}\circ\bar{\sigma}\circ\mathcal{T}_1(v,u).
\end{equation*}
In this construction, $\mathcal{T}_l(z)=A_lz+b_l$ represents some linear transformation with a matrix $A_l\in\mathbb{R}^{d_l\times d_{l-1}}$ and a vector $b_l\in\mathbb{R}^{d_l}$. $\bar{\sigma}:\mathbb{R}^{d_l}\mapsto \mathbb{R}^{d_{l+1}}$ applies the ReLU activiation funciton to each coordinate of the $d_l$-dimensional vector. 

Let $\mathcal G_n$ be this network class with weights and biases uniformly bounded by $R_{n}\ge1$. We allow for $L_n$ and $W_n$ that diverge with the sample size. Suppose $V$ has bounded support. With
$\log R_{n}+\log(W_nL_n)\lesssim\log n$, its uniform covering entropy has the leading order
\begin{equation}
 \log N(\eta,\mathcal G_n,\|\cdot\|_\infty)
 \lesssim W_n^2L_n^2\log(n/\eta).
 \label{eq:dda-dnn-network-entropy}
\end{equation}
see \citet[Lemma~A.1]{ChaiFanGu2026NDRE}. Set $F_g(y\mid v)=\Pr_\nu\{g(v,U)\le y\}$ and
$\mathcal F_n=\{F_g:g\in\mathcal G_n\}$. Coupling two generating functions with the same $U$ gives
\begin{equation}
 \begin{split}
 d_{2,\infty}(F_g,F_{g^\prime})^2
 \le\operatorname*{ess\,sup}_{v}
       \int|F_g(y\mid v)-F_{g^\prime}(y\mid v)|\,dy \le\|g-g^\prime\|_\infty.
 \end{split}
 \label{eq:dda-dnn-generator-transfer}
\end{equation}
Hence, a uniform $\eta^2$-cover of $\mathcal G_n$ provides an $\eta$-cover of the conditional distribution class in $d_{2,\infty}$. Combining \eqref{eq:dda-dnn-entropy-transfer}--\eqref{eq:dda-dnn-generator-transfer} yields condition~(iii) with $ S_n\asymp W_n^2L_n^2$ and $A_n\asymp n$.

One notable advantage of the DNN is its capability to explore certain low-dimensional compositional structures in the joint distribution, compared with the standard kernel smoothing or series expansion. Let $\mathcal C(d,r,\mathcal P)$ denote the hierarchical composition class. It is defined in a recursive way, and its depth-$r$ elements have the form
\[
 f(a)=g(f_1(a),\ldots,f_t(a)),\qquad
 f_i\in\mathcal C(d,r-1,\mathcal P),\quad (\beta,t)\in\mathcal P,
\]
where $g$ is $(\beta,C)$-smooth in the sense of \citet[Definition~2]{ChaiFanGu2026NDRE} and $\mathcal P\subset[1,\infty)\times\mathbb N$. With this smoothness restriction, write
$\gamma^*=\min_{(\beta,t)\in\mathcal P}\beta/t$.
If the true conditional quantile function belongs to $\mathcal C(d+1,r,\mathcal P)$ and the true CDF is uniformly Lipschitz in $y$, \citet[Corollary~4.2]{ChaiFanGu2026NDRE} give the following rates:
\[
 \epsilon_n\lesssim n^{-\gamma^*/(2\gamma^*+1)}(\log n)^a,
 \qquad
 c_n\lesssim n^{1/(2\gamma^*+1)}(\log n)^b,
\]
for some fixed $a,b\ge0$. The rate for $\epsilon_n$ is for the CDF in terms of metric $d_{L_2}$. It follows that our main restriction on the convergence rate becomes
\[
 c_n\epsilon_n^2
\lesssim n^{(1-2\gamma^*)/(2\gamma^*+1)}(\log n)^{b+2a},~~~
 c_n/\sqrt n
 \lesssim n^{(1-2\gamma^*)/\{2(2\gamma^*+1)\}}(\log n)^b.
\]
Both vanish when $\gamma^*>1/2$. This illustrates how low-dimensional compositional structure can satisfy the desired stochastic equicontinuity, even when the ambient dimension is large in complicated deep learning methods.

\section{Optimally weighted GMM and DA}

In this part, we substantiate the claims in the main text about the connection of our DA estimator and the optimal GMM estimator in \cite{chen2005MeasurementErrorModels}. In this appendix, we assume high-level assumptions with some smoothness for simplicity.

First, we provide a proof for the two-sum decomposition of the GMM objective function $\widehat{Q}_{GMM}$. Recall that
    $T:= \mathrm{Var}(\varsigma_0)$, $M := \mathrm{Cov}(s_0,\varsigma_0)\,\mathrm{Var}(\varsigma_0)^{-1}$, and
    \begin{align*}
        R &:= \mathrm{Var}(s_0) - \frac{1}{1+r}\,\mathrm{Cov}(s_0,\varsigma_0)\,\mathrm{Var}(\varsigma_0)^{-1}\,\mathrm{Cov}(s_0,\varsigma_0)^{\top}.
    \end{align*} and 
and let $\widehat T$, $\widehat{M}$ and $\widehat{R}$ be consistent estimators of $T$, $M$ and $R$, where $r$ is estimated by $n/N$.
\begin{proof}[Proof of \Cref{prop:gmm-decomposition}]
We begin with some notations. Let
\begin{align*}
    W_{O} :=  \begin{pmatrix}
            \mathrm{Var}(s_0) & - \mathrm{Cov}(s_0, \varsigma_0) \\
            -\mathrm{Cov}(s_0, \varsigma_0)^{\top} & (1+r) \mathrm{Var}(\varsigma_0)
        \end{pmatrix},
\end{align*}
where $\widehat{W}_O$ be an consistent estimator of $W_O$.
Then, $\widehat{Q}_{GMM}$ can be rewritten as follows.
\begin{align*}
    \widehat{Q}_{GMM}(\theta) =
    \begin{pmatrix}
        \mathbb{P}_n s_{\theta} \\
        \mathbb{Q}_N \varsigma_{\theta}(\cdot; \widehat{F}) - \mathbb{P}_n s_{\theta}
    \end{pmatrix}^{\top}
    \widehat{W}_{O}^{-1}
    \begin{pmatrix}
        \mathbb{P}_n s_{\theta} \\
        \mathbb{Q}_N \varsigma_{\theta}(\cdot; \widehat{F}) - \mathbb{P}_n s_{\theta}
    \end{pmatrix}.
\end{align*}
Note that block UDL decomposition (see e.g., \citet[Section 0.8.5]{horn2012} and \citet[Section 7.1]{bamieh2025linear}) gives 
\begin{align*}
    \widehat{W}_{O} = 
    \begin{pmatrix}
        I & - \frac{N}{n+N} \widehat{M} \\
        0 & I
    \end{pmatrix}
    \begin{pmatrix}
        \widehat{R} & 0 \\
        0 & \left(1+\frac{n}{N}\right)\widehat{T} \\
    \end{pmatrix}
    \begin{pmatrix}
        I & 0 \\
        -\frac{N}{n+N} \widehat{M}^{\top} & I
    \end{pmatrix}.
\end{align*}
and 
\begin{align*}
    \widehat{W}_{O}^{-1}
    =
    \begin{pmatrix}
        I & 0 \\
        \frac{N}{n+N}\widehat{M}^{\top}  & I
    \end{pmatrix}
    \begin{pmatrix}
        \widehat{R}^{-1} & 0 \\
        0 & \frac{N}{n+N}\widehat{T}^{-1} \\
    \end{pmatrix}
    \begin{pmatrix}
        I &\frac{N}{n+N}\widehat{M}  \\
        0 & I
    \end{pmatrix}.
\end{align*}
Therefore, we have
\begin{align*}
    \widehat{Q}_{GMM}(\theta) = 
    \widehat{Q}_{GMM, 1}(\theta) + \widehat{Q}_{GMM, 2}(\theta)
\end{align*}
where
\begin{align*}
    \widehat{Q}_{GMM, 1}(\theta) 
    & = 
    \left\| \mathbb{P}_n s_{\theta} + \frac{N}{n+N} \widehat{M} \left( \mathbb{Q}_N \varsigma_{\theta}(\cdot; \widehat{F}) - \mathbb{P}_n s_{\theta}\right)\right\|_{\widehat{R}^{-1}}^2 \\
    & = 
    \left\| \left(I - \frac{N}{n+N} \widehat{M}\right) \mathbb{P}_n s_{\theta} + \frac{N}{n+N} \widehat{M} \mathbb{Q}_N \varsigma_{\theta}(\cdot; \widehat{F}) \right\|_{\widehat{R}^{-1}}^2,
\end{align*}
and
\begin{align*}
    \widehat{Q}_{GMM, 2}(\theta) = \frac{N}{n+N}\left\| \mathbb{Q}_N \varsigma_{\theta}(\cdot; \widehat{F}) - \mathbb{P}_n s_{\theta} \right\|_{\widehat{T}^{-1}}^2.
\end{align*}
This concludes the proof.
\end{proof}

Following \cite{chen2005MeasurementErrorModels}, we can show that the GMM estimator follows asymptotic normality under some mild regularity conditions. Here, we carefully examine the second term in the optimally weighted GMM objective function. Proposition \ref{prop:gmm-second-term-negligible} is new, to the best of our knowledge. The result may be of independent of interest for any root-$n$ consistent estimator $\widehat\theta$. For simplicity, we assume the following assumptions, which correspond to Conditions (A.1) and (A.2) in the proof of Theorem 2 in \cite{chen2005MeasurementErrorModels}.

\begin{assumption} \label{ass:score-AL-hessian}
    Assume that $s_{\theta}$ and $\varsigma_{\theta}$ are continuously differentiable with respect to $\theta$, and $\widehat{F}$ satisfies the following.
    \begin{gather*}
        \sup_{\|\theta - \theta_0\|=o(1)} \| \mathbb{P}_n \partial_{\theta} s_{\theta} - \mathbb{E}[\partial_{\theta} s_{0}(Z_i)] \| = o_p(1) \\
        \sup_{\|\theta - \theta_0\|=o(1)} \| \mathbb{Q}_N \partial_{\theta} \varsigma_{\theta}(\cdot, \widehat{F}) - \mathbb{E}[\partial_{\theta} \varsigma_{0}(V_i, F_0)] \| = o_p(1)
    \end{gather*}
    and let $H = \mathbb{E}[\partial_{\theta} s_{\theta_0}(Z_i)] = \mathbb{E}[\partial_{\theta} \varsigma_{\theta_0}(V_i, F_0)]$.
\end{assumption}

\begin{assumption} \label{ass:varsigma-AL}
\begin{align*}
    \mathbb{Q}_N \varsigma_{0}(\cdot, \widehat{F}) = \mathbb{P}_n (s_{0} - \varsigma_{0}(\cdot, F_0)) + \mathbb{Q}_N \varsigma_{0}(\cdot, F_0) + o_p(n^{-1/2}).
\end{align*}
where $r = \lim_{n \to \infty} (n/N)$.
\end{assumption}

\begin{proposition} \label{prop:gmm-second-term-negligible}
Under Assumptions \ref{Assump:Sampling}, \ref{Assump:ID}, \ref{ass:score-AL-hessian} and \ref{ass:varsigma-AL}, we have
\begin{align*}
   \partial_{\theta} \widehat{Q}_{GMM, 2}(\widehat{\theta}) = o_p(n^{-1/2})
\end{align*}
for any estimator $\widehat{\theta}$ such that $\widehat{\theta} - \theta_0 = O_p(n^{-1/2})$.

\end{proposition} 

\begin{proof}[Proof of \Cref{prop:gmm-second-term-negligible}]
In the following proof, we handle $(n+N)/N\times  \partial_{\theta}\widehat{Q}_{GMM,2}$ as $n/N \to r$. 
Note that 
    \begin{align*}
         \left(\frac{n+N}{N}\right) \frac{1}{2}\sqrt{n} \partial_{\theta} \widehat{Q}_{GMM, 2}(\widehat{\theta}) 
         & = 
         \left(\mathbb{Q}_N\partial_{\theta}\varsigma_{\widehat{\theta}}(\cdot;\widehat{F})
    -\widehat{P}_n \partial_{\theta} s_{\widehat{\theta}}\right)^{\top} \widehat{T}^{-1} \sqrt{n} \left(\mathbb{Q}_N \varsigma_{\widehat{\theta}}(\cdot;\widehat{F})
    - \mathbb{P}_n s_{\widehat{\theta}}\right) \\
    & = \left(\mathbb{Q}_N  \partial_{\theta}\varsigma_{\widehat{\theta}}(\cdot;\widehat{F})
    - \mathbb{P}_n \partial_{\theta} s_{\widetilde{\theta}} \right)^{\top} \widehat{T}^{-1} \left(\mathbb{Q}_N  \partial_{\theta}\varsigma_{\widetilde{\theta}}(\cdot;\widehat{F})
    - \mathbb{P}_n \partial_{\theta} s_{\widetilde{\theta}}\right) \sqrt{n}(\widehat{\theta} - \theta_0) \\
    & \quad +
    \left(\mathbb{Q}_N \partial_{\theta}\varsigma_{\widehat{\theta}}(\cdot;\widehat{F})
    - \mathbb{P}_n \partial_{\theta} s_{\widehat{\theta}} \right)^{\top} \widehat{T}^{-1} \sqrt{n} \left(\mathbb{Q}_N \varsigma_{0}(\cdot;\widehat{F})
    - \mathbb{P}_n s_{0} \right),
    \end{align*}
    where $\widetilde{\theta}$ is between $\widehat{\theta}$ and $\theta_0$. 
First, Assumption \ref{ass:score-AL-hessian} imply that
\begin{align*}
    \left(\mathbb{Q}_N \partial_{\theta}\varsigma_{\widehat{\theta}}(V_i;\widehat{F})
    - \mathbb{P}_n \partial_{\theta} s_{\widehat{\theta}}\right) \xrightarrow{p} H - H = 0.
\end{align*}
Second, Assumptions \ref{Assump:Sampling}, \ref{Assump:ID} and \ref{ass:varsigma-AL} imply that
\begin{align*}
     \sqrt{n} \left(\mathbb{Q}_N  \varsigma_{0}(\cdot;\widehat{F})
    -\mathbb{P}_n s_{0}\right)
    = O_p(1).
\end{align*}  
Therefore, $\frac{1}{2}\sqrt{n} \partial_{\theta} \widehat{Q}_{GMM, 2}(\widehat{\theta}) = o_p(1)$ as long as $\widehat{\theta} - \theta_0 = O_p(n^{-1/2})$, which concludes the proof. 
\end{proof}
\Cref{prop:gmm-second-term-negligible} implies that our DA estimator approximately solves the first order condition in the optimal GMM problem, because of $\partial_{\theta} \widehat{Q}_{GMM, 1}(\widehat{\theta}_{DA}) = o_p(n^{-1/2})$ and $\partial_{\theta} \widehat{Q}_{GMM, 2}(\widehat{\theta}_{DA}) = o_p(n^{-1/2})$. The same proposition also implies that the optimal GMM estimator $\widehat{\theta}_{GMM}$ solves our DA problem approximately under mild regularity conditions because $\widehat{Q}_{GMM, 1}(\theta) \approx \|\partial_{\theta} \widehat{Q}_{DA}(\theta)\|_{\widehat{R}^{-1}}^2$. As discussed in \cite{chen2005MeasurementErrorModels}, the optimal GMM estimator $\widehat{\theta}_{GMM}$ satisfies $\sqrt{n}$-the consistency and asymptotic normality under condition therein. This implies $\partial_{\theta}\widehat{Q}_{GMM}(\widehat{\theta}_{GMM}) = o_p(n^{-1/2})$ and $\partial_{\theta}\widehat{Q}_{GMM, 2}(\widehat{\theta}_{GMM}) = o_p(n^{-1/2})$. Therefore, $\partial_{\theta} \widehat{Q}_{GMM, 1}(\widehat{\theta}_{GMM}) = o_p(n^{-1/2})$ and $\partial_{\theta} \widehat{Q}_{DA}(\widehat{\theta}_{GMM}) = o_p(n^{-1/2})$.

\section{Computation Algorithm}

In this part, we describe the computational procedures of the cross-fitted version of our estimation methods, and the corresponding multiplier bootstrap procedure.

For completeness, we recall the estimation procedure using the full sample. Let
\[
\mathcal{I}_L = \{1,\dotsc,n\},
\quad
\mathcal{I}_U = \{n+1,\dotsc,n+N\},
\quad
r_n = \frac{n}{N}, \quad a_n = \frac{r_n}{1+r_n}, \quad b_n = \frac{1}{1+r_n},
\]
where $\mathcal{I}_L$ and $\mathcal{I}_U$ index the labeled and unlabeled data, respectively. The DA and DDA estimators can be written as follows.
\begin{small}
\begin{align*}
    \widehat{\theta}_{DA}
    &=
    \argmin_{\theta}
    \Bigg\{
        \frac{a_n}{|\mathcal{I}_L|}
        \sum_{i\in\mathcal{I}_L}
        \ell_{\theta}(X_i,Y_i)
        +
        \frac{b_n}{|\mathcal{I}_U|}
        \sum_{j \in\mathcal{I}_U}
        \mu_{\theta}(X_j, \widehat{Y}_j ;\widehat{F})
    \Bigg\},
    \\
    \widehat{\theta}_{DDA}
    &=
    \argmin_{\theta\in\Theta}
    \Bigg\{
        \frac{1}{|\mathcal{I}_L|}
        \sum_{i\in\mathcal{I}_L}
        \ell_{\theta}(X_i,Y_i)
        +
        b_n
        \Bigg[
            \frac{1}{|\mathcal{I}_U|}
            \sum_{j\in\mathcal{I}_U}
             \mu_{\theta}(X_j, \widehat{Y}_j ;\widehat{F})
            -
            \frac{1}{|\mathcal{I}_L|}
            \sum_{i\in\mathcal{I}_L}
            \mu_{\theta}(X_i, \widehat{Y}_i ;\widehat{F})
        \Bigg]
    \Bigg\}.
\end{align*}
\end{small}

\subsection{Cross-fitted Semiparametric Estimation.}
Here, we present the algorithm for cross-fitted estimation. We randomly partition the labeled data into $K$ folds with index sets
$\mathcal{I}_{L,1},\dotsc,\mathcal{I}_{L,K}$.
For each $k=1,\dotsc,K$, we estimate the conditional distribution
$\widehat{F}_{-k}(\cdot\mid X,\widehat{Y})$
using the labeled data outside fold $k$, and compute DA or DDA estimator using the fold $k$ and unlabeled data.
\begin{align*}
 \widehat{\theta}_{DA, k}
    &=
    \argmin_{\theta\in\Theta}
    \Bigg\{
        \frac{a_n}{|\mathcal{I}_{L,k}|}
        \sum_{i\in\mathcal{I}_{L,k}}
        \ell_{\theta}(X_i,Y_i)
        +
            \frac{b_n}{|\mathcal{I}_U|}
            \sum_{j\in\mathcal{I}_U} 
            \mu_{\theta}(X_j, \widehat{Y}_j; \widehat{F}_{-k})
    \Bigg\}.
\\
    \widehat{\theta}_{DDA, k}
    &=
    \argmin_{\theta\in\Theta}
    \Bigg\{
        \frac{1}{|\mathcal{I}_{L,k}|}
        \sum_{i\in\mathcal{I}_{L,k}}
        \ell_{\theta}(X_i,Y_i)
        \\
    &\qquad
        +
        b_n
        \Bigg[
            \frac{1}{|\mathcal{I}_U|}
            \sum_{j \in\mathcal{I}_U}
           \mu_{\theta}(X_j, \widehat{Y}_j; \widehat{F}_{-k})
            -
            \frac{1}{|\mathcal{I}_{L,k}|}
            \sum_{i\in\mathcal{I}_{L,k}}
            \mu_{\theta}(X_i, \widehat{Y}_i; \widehat{F}_{-k})
        \Bigg]
    \Bigg\}.
\end{align*}
The cross-fitted estimators of $DA$ and $DDA$ are given by
\[
    \widehat{\theta}_{DA}
    =
    \frac{1}{K}
    \sum_{k=1}^{K}
    \widehat{\theta}_{DA, k}, \quad \widehat{\theta}_{DDA}
    =
    \frac{1}{K}
    \sum_{k=1}^{K}
    \widehat{\theta}_{DDA, k}.
\]

\subsection{Cross-fitted Multiplier Bootstrap.}
Here, we describe the computational detail of a cross-fitted bootstrap procedure, in line with the general methods from \cite{LinHan2026BootstrapDML}. We generate bootstrap realizations of
$\widetilde{T}_n^*:=T_n^*/\sqrt{n}$
while holding the folds and estimated conditional distributions fixed. Let $\widehat{H}_k$ denote the estimate of Hessian matrix $H$ that we obtained when we compute
$\widehat{\theta}_k$. Recall the following notation:
\[
    \widehat s_i:=s_{\widehat\theta}(X_i,Y_i),
    \qquad
    \widehat \varsigma_{i, k}:=\varsigma_{\widehat\theta}(V_i;\widehat F_{-k}), \qquad \widehat{v}_{i, k} = \widehat{s}_i - b_n \widehat{\varsigma}_{i, k}.
\]

To approximate the conditional score expectation $\widehat \varsigma_{i, k}$, we generate $J$ draws
\[
    Y_{-k,i}^{(s)}
    \sim
    \widehat{F}_{-k}(\cdot\mid X_i,\widehat{Y}_i),
    \qquad s=1,\dotsc,S,
\]
for each $k=1,\dotsc,K$ and
$i\in\mathcal{I}_{L,k}\cup\mathcal{I}_U$.
Use the approximation
\[
    \widehat \varsigma_{i, k} = \mathbb{E}_{\widehat{F}_{-k}}
    \bigl[
        s_{\widehat{\theta}}(X_i,Y_i)
        \mid X_i,\widehat{Y}_i
    \bigr]
    \approx
    \frac{1}{S}
    \sum_{s=1}^{S}
    s_{\widehat{\theta}}(X_i,Y_{-k,i}^{(s)}).
\]
We generate these outcomes once and reuse them across all bootstrap replications. For each replication $b=1,\dotsc,B$, we draw the weights
$\{\xi_i^{(b)}\}_{i\in\mathcal{I}_L}$ and
$\{\zeta_j^{(b)}\}_{j\in\mathcal{I}_U}$ once and use them across all folds.

Define
\[
    \bar{\xi}_k^{(b)}
    =
    \frac{1}{|\mathcal{I}_{L,k}|}
    \sum_{i\in\mathcal{I}_{L,k}}
    \xi_i^{(b)},
    \qquad
    \bar{\zeta}^{(b)}
    =
    \frac{1}{|\mathcal{I}_U|}
    \sum_{j\in\mathcal{I}_U}
    \zeta_j^{(b)}.
\]
For each fold $k$, we compute
\begin{align*}
    \widetilde{T}_{b,k}^*
    =
    -\widehat{H}_k^{-1}
    \Bigg\{&
        \frac{1}{|\mathcal{I}_{L,k}|}
        \sum_{i\in\mathcal{I}_{L,k}}
        \bigl(\xi_i^{(b)}-\bar{\xi}_k^{(b)}\bigr)
        \widehat{v}_{i, k}
    +
            \frac{b_n}{|\mathcal{I}_U|}
            \sum_{j\in\mathcal{I}_U}
            \bigl(\zeta_j^{(b)}-\bar{\zeta}^{(b)}\bigr)
            \widehat{\varsigma}_{j, k}
    \Bigg\}.
\end{align*}
The $b$-th bootstrap realization is
\[
    \widetilde{T}_b^*
    =
    \frac{1}{K}
    \sum_{k=1}^{K}
    \widetilde{T}_{b,k}^*.
\]
The collection of $\{\widetilde{T}_b^*\}_{b=1}^{B}$ will be used for inference.

\end{document}